\documentclass[showpacs,aps,prd,reprint,superscriptaddress,nofootinbib,longbibliography]{revtex4-1}
\usepackage[colorlinks=true, pdfstartview=FitV, linkcolor=magenta,citecolor=blue, urlcolor=magenta,
bookmarks=true, bookmarksnumbered=true, breaklinks]{hyperref}
\usepackage[dvipdfmx]{graphicx}

\usepackage{amsmath,amssymb,bm,color,longtable,mathrsfs,amsfonts,slashed}

\newcommand{\Slash}[1]{{\ooalign{\hfil/\hfil\crcr$#1$}}}

\begin{document}
\begin{flushright}
\end{flushright}

\title{Kondo effect driven by chirality imbalance}

\author{Daiki~Suenaga}
\email[]{\tt suenaga@mail.ccnu.edu.cn}
\affiliation{Key Laboratory of Quark and Lepton Physics (MOE) and Institute of Particle Physics, Central China Normal University, Wuhan 430079, China}

\author{Kei~Suzuki}
\email[]{{\tt k.suzuki.2010@th.phys.titech.ac.jp}}
\affiliation{Advanced Science Research Center, Japan Atomic Energy Agency (JAEA), Tokai 319-1195, Japan}

\author{Yasufumi~Araki}
\email[]{\tt araki.yasufumi@jaea.go.jp}
\affiliation{Advanced Science Research Center, Japan Atomic Energy Agency (JAEA), Tokai 319-1195, Japan}

\author{Shigehiro~Yasui}
\email[]{\tt yasuis@keio.jp}
\affiliation{Research and Education Center for Natural Sciences,\\ Keio University, Hiyoshi 4-1-1, Yokohama, Kanagawa 223-8521, Japan}

\date{\today}

\begin{abstract}
We propose a novel mechanism of the Kondo effect driven by a chirality imbalance (or chiral chemical potential) of relativistic light fermions.
This effect is realized by the mixing between a right- or left-handed fermion and a heavy impurity in the chirality imbalanced matter even at zero density.
This is different from the usual Kondo effect induced by finite density.
We derive the Kondo effect from both a perturbative calculation and a mean-field approach.
We also discuss the temperature dependence of the Kondo effect.
The Kondo effect at nonzero chiral chemical potential can be tested by future lattice simulations.
\end{abstract}


\maketitle

\section{Introduction}
The  Kondo effect~\cite{Kondo:1964,Hewson,Yosida,Yamada,coleman_2015} is  known as a phenomenon which occurs in metal including heavy impurities.
It leads to drastic modifications of the transport properties of conducting (or itinerant) electrons at low temperature.
While, in the conventional case, itinerant electrons are treated as nonrelativistic fermions,
recent studies show that the Kondo effect can be realized also in systems with relativistic fermions. 

One example of the Kondo effect realized in the relativistic system is the {\it isospin Kondo effect}. This effect can be induced near the Fermi surface of nucleons with a heavy hadron such as $\Sigma_c^{(*)}$ or $\bar{D}^{(*)}$ existing as an impurity, where the non-Abelian $SU(2)$ interaction between the light nucleon and the heavy hadron is supplied by an isospin exchange~\cite{Yasui:2013xr,Yasui:2016ngy,Yasui:2016hlz,Yasui:2019ogk}. In the context of quantum chromodynamics (QCD) in which the non-Abelian $SU(3)$ interaction is governed by the color interaction mediated by gluons, the so-called {\it QCD Kondo effect} which may be realized in quark matter composed of up and down (and also often strange) quarks with a heavy (charm or bottom) quark, has been studied in the literatures~\cite{Yasui:2013xr,Hattori:2015hka,Ozaki:2015sya,Yasui:2016svc,Yasui:2016yet,Kanazawa:2016ihl,Kimura:2016zyv,Yasui:2017izi,Suzuki:2017gde,Yasui:2017bey,Kimura:2018vxj,Macias:2019vbl,Hattori:2019zig,Suenaga:2019car}.
Moreover, in the context of electron systems with Dirac or Weyl dispersion in solid states, called Dirac/Weyl semimetals, it has been seen that the vanishing density of states around the Dirac or Weyl points leads to an anomalous Kondo screening behavior, distinct from normal metals~\cite{Principi:2015,Yanagisawa:2015conf,Yanagisawa:2015,Mitchell:2015,Sun:2015,Feng:2016,Kanazawa:2016ihl,Lai:2018,Ok:2017,PhysRevB.97.045148,PhysRevB.98.075110,PhysRevB.99.115109,KIM2019236,Grefe:2019}.

In relativistic massless fermions, one of the interesting characteristics is their chirality {\it i.e.} the left-handed and right-handed degrees of freedoms. In this paper, we propose a novel type of Kondo effect: the Kondo effect driven by a chirality imbalance (or chiral chemical potential $\mu_5$).
This is similar to the ``usual" Kondo effect induced on the Fermi surface but slightly different in the sense that it occurs even at zero chemical potential, $\mu = 0$. We particularly study the Kondo effect at finite $\mu_5$ perturbatively and non-perturbatively: The former is accomplished by the renormalization group (RG) analysis at one loop, and the latter is by the mean-field analysis. In the present work, we do not take into account the effects from interactions between two light fermions such as chiral condensate in order to focus on the Kondo effect in a transparent way.

To investigate systems with $\mu_5$ will give a motivation for Monte Carlo (lattice) simulations of strongly correlated quantum systems such as the Kondo effect and quark-gluon dynamics, which is one of the promising tools to nonperturbatively study them.
While Monte Carlo simulations with a finite chemical potential $\mu$ suffer from the sign problem, at finite chiral chemical potential $\mu_5$, the sign problem is absent~\cite{Fukushima:2008xe} (also see Refs.~\cite{Yamamoto:2011gk,Yamamoto:2011ks,Braguta:2015zta,Braguta:2015owi,Astrakhantsev:2019wnp}).
Therefore, when the Kondo effects are induced by finite $\mu_5$, we expect that Monte Carlo simulations with $\mu_5$ would be promising for measuring the Kondo effect.

In the context of QCD, a chirality imbalance might be realized in the heavy-ion collision (HIC) experiments.
Arguments on its possibility have a long history~\cite{Morley:1983wr}, and there are some scenarios leading to local parity violation, such as the sphaleron transition~\cite{McLerran:1990de,Kharzeev:2001ev,Kharzeev:2007jp}, the parallel color electric and magnetic fields (or the Glasma)~\cite{Kharzeev:2001ev,Lappi:2006fp}, and disoriented pseudoscalar condensates~\cite{Kharzeev:1998kz,Kharzeev:1999cz}.
At early state of HIC, heavy quarks are also produced by hard processes mediated by gluons from nucleon-nucleon scattering at high energy.\footnote{These heavy quarks can play a role of heavy impurities without satisfying any chemical equilibrium conditions, when we focus on the short time scale in which  the weak decay of the heavy quarks does not take place. The Kondo effect can evolve regardless of any chemical equilibriums on the heavy quarks.}
Thus HIC is expected to be a possible environment to study the Kondo effect with chirality imbalance.\footnote{The realization of chiral chemical potential large enough in experiments is an open question. For example, chiral charges are not conserved because of the existence of the quantum anomaly.}

Our analyses can also be extended to Dirac or Weyl semimetals with energy splitting among Dirac or Weyl cones in electronic band structure, such as in Weyl semimetals with broken inversion symmetry~\cite{Zyuzin:2012a,Zyuzin:2012b,Vazifeh:2013}.
Such an effect may be reproduced as well by Zeeman splitting of spin-degenerate Dirac cones in topological Dirac semimetals~\cite{Burkov:2016}, such as Cd${}_{3}$As${}_{2}$~\cite{Wang:2013,Neupane:2014}.

This paper is organized as follows.
In Sec.~\ref{Sec_2}, we consider the Kondo effect at finite $\mu_5$ from an effective Lagrangian and a perturbation calculation.
In Sec.~\ref{Sec_3}, to study the Kondo effect in the nonperturbative region, we formalize a mean field approach, and show the phase diagram of the Kondo effect on the plane of temperature and $\mu_5$.
Section~\ref{Sec_4} is devoted to our conclusion and outlook.

\section{Perturbative approach} \label{Sec_2}

In this section, we show the emergence of the Kondo effect at finite $\mu_5$ within a perturbative scheme, which can be signaled by existence of a Landau pole in the renormalization group (RG) flow for the effective coupling between a light fermion and a heavy fermion~\cite{Anderson:1970}.\footnote{A perturbative calculation of the QCD Kondo effect at finite $\mu_5$ was also done in an early work by Ozaki and Itakura (unpublished).}

\begin{figure*}[t!]
\centering
\includegraphics*[scale=0.44]{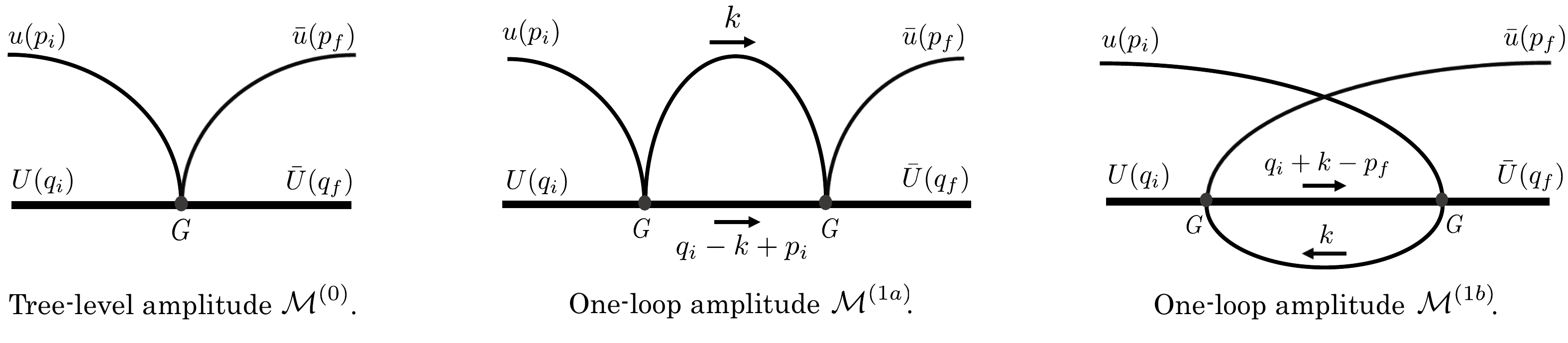}
\caption{The diagrammatical picture of the scattering amplitude between the light fermion and the heavy fermion in Eqs.~(\ref{PZeroth}),~(\ref{POneLoop1}) and~(\ref{POneLoop2}).}
\label{fig:Amplitude}
\end{figure*}

We start our discussion by the following Lagrangian to describe a scattering between a light fermion and a heavy fermion:
\begin{eqnarray}
{\cal L} &=& \bar{\psi} (i\Slash{\partial} + \mu_5 \gamma_0 \gamma_5) \psi + \bar{\Psi}(i\Slash{\partial}-M_Q)\Psi \nonumber\\
&& +G(\bar{\psi}t^a\gamma^\mu\psi)(\bar{\Psi}t^a\gamma_\mu\Psi)\ , \label{PLagrangian}
\end{eqnarray}
in which $\psi$ and $\Psi$ denote the light-fermion and heavy-fermion fields, respectively. $\mu_5$ is the chiral chemical potential and $M_Q$ is the heavy-fermion mass whose value is significantly larger than the typical scale of the theory.
$t^a$ with an index $a=1,\dots,N^{2}-1$ is the generator of the $SU(N)$ group characterizing a non-Abelian interaction.
In terms of the interaction manner between the light fermion and the heavy fermion, we have employed a vector-type contact interaction.\footnote{In the context of QCD, the interaction term in Eq.~(\ref{PLagrangian}) can be motivated by a one-gluon exchange interaction between the light quark and the heavy quark with a large Debye mass, as demonstrated in Ref.~\cite{Hattori:2015hka}. This is confirmed at one-loop calculation for $\mu_5$ without ordinary chemical potential $\mu$.}
$G>0$ is the coupling constant.
We notice that, in this section, we introduce the heavy-fermion field ($\Psi$) as a Dirac spinor which includes an anti-particle as well as a particle component.
However, later, we will take a limit of $M_Q\to\infty$ to describe the emergence of the Kondo effect in more transparent way.

The scattering amplitude between the light fermion and the heavy fermion up to one loop is of the form
\begin{eqnarray}
{\cal M}= {\cal M}^{(0)}+ {\cal M}^{(1)}\ , \label{M0PlusM1}
\end{eqnarray}
where ${\cal M}^{(0)}$ and ${\cal M}^{(1)}$ are the amplitude at tree level and at one-loop level, respectively. 
Explicitly, ${\cal M}^{(0)}$ and ${\cal M}^{(1)}$ are obtained as
\begin{eqnarray}
{\cal M}^{(0)} &=& G \bar{u}(p_f)t^a\gamma^\mu u(p_i) \bar{U}(q_f)t^a\gamma_\mu U(q_i) \ ,\label{PZeroth}
\end{eqnarray}
and
\begin{eqnarray}
{\cal M}^{(1)} = {\cal M}^{(1a)} + {\cal M}^{(1b)}\ , \label{POneLoop}
\end{eqnarray}
with
\begin{eqnarray}
{\cal M}^{(1a)} &=& -G^2 T\sum_{n}\int\frac{d^3k}{(2\pi)^3} \bar{u}(p_f) t^a\gamma^\mu S_l(k)t^b\gamma^\nu u(p_i) \nonumber\\
&& \times\bar{U}(q_f)t^a\gamma_\mu S_h(q_i-k+p_i)t^b\gamma_\nu U(q_i)\ , \label{POneLoop1}
\end{eqnarray}
and
\begin{eqnarray}
{\cal M}^{(1b)} &=&-G^2 T\sum_{n}\int\frac{d^3k}{(2\pi)^3}\bar{u}(p_f)t^a\gamma^\mu S_l(k) t^b\gamma^\nu u(p_i) \nonumber\\ \nonumber\\  
&& \times\bar{U}(q_f)t^b\gamma_\nu S_h(q_i+k-p_f)t^a\gamma_\mu U(q_i) \ , \label{POneLoop2}
\end{eqnarray}
respectively, which are diagrammatically indicated in Fig.~\ref{fig:Amplitude}.
$u(p)$ and $U(q)$ are the Dirac wavefunctions for the light and heavy fermions, respectively, with $p=p_{i}$ ($p_{f}$) and $q=q_{i}$  ($q_{f}$) the initial (final) momenta.
In Eq.~(\ref{POneLoop}), we have employed the imaginary-time formalism to take into account the finite temperature effect, so that the propagators $S_l(k)$ and $S_h(k)$ take the form of
\begin{eqnarray}
S_l(k) = \sum_{\epsilon_5=\pm}P_{\epsilon_5}\tilde{\Delta}_l\big(i(\omega_{n}-i\epsilon_5\mu_5)\big) \ ,  \label{SlMatsubara}
\end{eqnarray}
and
\begin{eqnarray}
S_h(k) = \tilde{\Delta}_h(i\omega_{n})\ ,
\end{eqnarray}
with
\begin{eqnarray}
\tilde{\Delta}_l\big(i(\omega_{n}-i\epsilon_5\mu_5)\big) = -\frac{i(-\omega_{n}+i\epsilon_5\mu_5)\gamma_0+\vec{k}\cdot\vec{\gamma}}{(\omega_{n}-i\epsilon_5\mu_5)^2+|\vec{k}|^2}\ ,  \label{DeltaLDef}
\end{eqnarray}
and
\begin{eqnarray}
\tilde{\Delta}_h(i\omega_{n}) = -\frac{-i\omega_{n}\gamma_0+\vec{k}\cdot\vec{\gamma}-M_Q}{\omega_{n}^2+|\vec{k}|^2 +M_Q^2}\ , \label{DeltaHDef}
\end{eqnarray}
where $\vec{\gamma} \equiv (\gamma^1, \gamma^2, \gamma^3)$ is the spatial components of the Dirac gamma matrices.
In these expressions, $P_\pm=(1\pm\gamma_5)/2$ is the right-handed or left-handed projection operator, and the Matsubara frequency is $\omega_{n}=(2n+1)\pi T$ ($n=0,\pm1,\pm2,\cdots$). The detailed calculation of the one loops in Eqs.~(\ref{POneLoop1}) and~(\ref{POneLoop2}) within the imaginary-time formalism is provided in Appendix~\ref{sec:MatsubaraSum}.

Before showing the results of Eq.~(\ref{M0PlusM1}), we notice some important points about the fermion wavefunctions $u(p)$ ($\bar{u}(p)$) or $U(q)$ ($\bar{U}(q)$). First, in terms of the light-fermion wavefunction, it is useful to separate the light-fermion transition part in Eq.~(\ref{M0PlusM1}) into the right-handed and left-handed ones by defining $u_R=P_+ u$ and $u_L=P_- u$, since the Lagrangian~(\ref{PLagrangian}) preserves the axial current.

Next, in terms of the heavy-fermion wavefunction, as is well known, the free Dirac spinor can be decomposed into 
\begin{eqnarray}
U(q) &=& \Lambda_+U(q)  + \Lambda_-U(q)  \nonumber\\
&=& U_+(q) + U_-(q)\ , \label{UDecompose}
\end{eqnarray}
with $U_\pm(q) \equiv \Lambda_\pm U(q)$, by defining the projection operator with respect to the positive-energy ($+$) and negative-energy ($-$) solutions of the Dirac equation:
\begin{eqnarray}
\Lambda_\pm \equiv \frac{ M_Q\pm(q_0\gamma_0-\vec{q}\cdot\vec{\gamma})}{2M_Q} ,
\end{eqnarray}
with $q_0=\sqrt{|\vec{q}|^2+M_Q^2}$.
When we measure the energy of the fermion from $M_Q$ as in the non-relativistic system, {\it i.e.}, by shifting the energy of the positive-energy and negative-energy components commonly as $q_0 \to q_0-M_Q$, we need to cost at least $2M_Q$ for the excitation of the negative-energy component, which can be ignored in the limit of $M_Q\to \infty$. Therefore, when we consider such a situation, we can drop $U_-(q)$ in Eq.~(\ref{UDecompose}), and replace $U(q)$ by $U_+(q)$.

By taking the above arguments into account, the tree-level amplitude in Eq.~(\ref{PZeroth}) can be reduced to
\begin{eqnarray}
{\cal M}^{(0)} &=& G \, \bar{u}_R(p_f)t^a\gamma_0 u_R(p_i) \bar{U}_+(q_f)t^a U_+(q_i) \nonumber\\
&+& G \, \bar{u}_L(p_f)t^a\gamma_0 u_L(p_i) \bar{U}_+(q_f)t^a U_+(q_i)\ , \label{MZeroComp}
\end{eqnarray}
in which we have used a fact of $\bar{U}_+(q_f)t^a\vec{\gamma} U_+(q_i) =0$ with $M_Q\to\infty$. The one-loop amplitude ${\cal M}^{(1)}$ in Eq.~(\ref{POneLoop}) is calculated in detail in Appendix~\ref{sec:MatsubaraSum}. According to Eq.~(\ref{MOneCompApp}), the resulting ${\cal M}^{(1)}$ is of the form
\begin{eqnarray}
{\cal M}^{(1)} &\approx&  \frac{G^2}{2}\frac{N\rho_0}{2}\int_{-\mu_5}^\infty dE\frac{1-\tilde{f}_{\beta}(E)}{E} \nonumber\\
&& \times\bar{u}_R(p_f)t^a \gamma_0 u_R(p_i)\bar{U}_+(q_f)t^a U_+(q_i) \nonumber\\
&&+ \frac{G^2}{2}\int\frac{d^3k}{(2\pi)^3} \frac{1}{\mu_5-|\vec{k}|} \nonumber\\
&& \times\bar{u}_L(p_f)t^at^b\gamma_0u_L(p_i) \bar{U}_+(q_f)t^bt^aU_+(q_i)\ ,\label{MOneComp}
\end{eqnarray} 
in the limit of $M_Q\to\infty$ with the initial- and final-state light fermions inhabiting the ``Fermi surface'', {\it i.e.} the initial- and finial-state light fermions satisfy the kinematics of $(p^0, |\vec{p}|) = (0,\mu_5)$ for the right-handed fermion while $(p^0, |\vec{p}|) = (2\mu_5,\mu_5)$ for the left-handed fermion ($p^\mu$  stands for $p^\mu_i$ and $p^\mu_f$ collectively), due to the Dirac equation.
$\tilde{f}_{\beta}(E)$ 
 is the Fermi distribution function, $\tilde{f}_{\beta}(E) = 1/({e}^{\beta E} + 1)$ with inverse temperature $\beta=1/T$. In obtaining Eq.~(\ref{MOneComp}), we notice that the density of states at Fermi surface $\rho_0 = \mu_5^2/(2\pi^2)$ is employed since we assume implicitly a hierarchy of $M_Q(\to\infty) \gg \mu_5 \gg T$.

From the above considerations, it turns out that Eqs.~(\ref{MZeroComp}) and~(\ref{MOneComp}) lead to the RG equation~\cite{Anderson:1970} as
\begin{eqnarray}
\Lambda\frac{dG(\Lambda)}{d\Lambda}
= -
\frac{N\rho_0G^2(\Lambda)}{4}
(1-\tilde{f}_{\beta}(\Lambda))\ , \label{RGEMu5}
\end{eqnarray}
for the coupling $G(\Lambda)$ of only the right-handed fermion, where the effective coupling $G(\Lambda)$ depends on the energy scale $\Lambda$ measured from the Fermi surface.
Alternatively, the RG equation~(\ref{RGEMu5}) can be converted into the dimensionless one as
\begin{eqnarray}
\bar{\Lambda}\frac{d\bar{G}(\bar{\Lambda})}{d\bar{\Lambda}}
= -\frac{N\bar{G}(\bar{\Lambda})^2}{8\pi^2}
(1-\tilde{f}_{\bar{\beta}}(\bar{\Lambda})),
\label{RGEMu52}
\end{eqnarray}
by defining $\bar{\Lambda}=\Lambda/\mu_5$, $\bar{G}(\bar{\Lambda}) = G(\Lambda)\mu_5^2$, and $\bar{\beta}=\beta\mu_{5}$ ($\bar{T}=T/\mu_5$). 
We comment that Eq.~\eqref{RGEMu52} is reduced to the simple form,
\begin{eqnarray}
   \bar{\Lambda}\frac{d\bar{G}(\bar{\Lambda})}{d\bar{\Lambda}} = -\frac{N\bar{G}(\bar{\Lambda})^2}{8\pi^2},
\label{RGEMu52_0}
\end{eqnarray}
at $\bar{T}=0$.

\begin{figure}[t!]
\centering
\includegraphics*[scale=0.7]{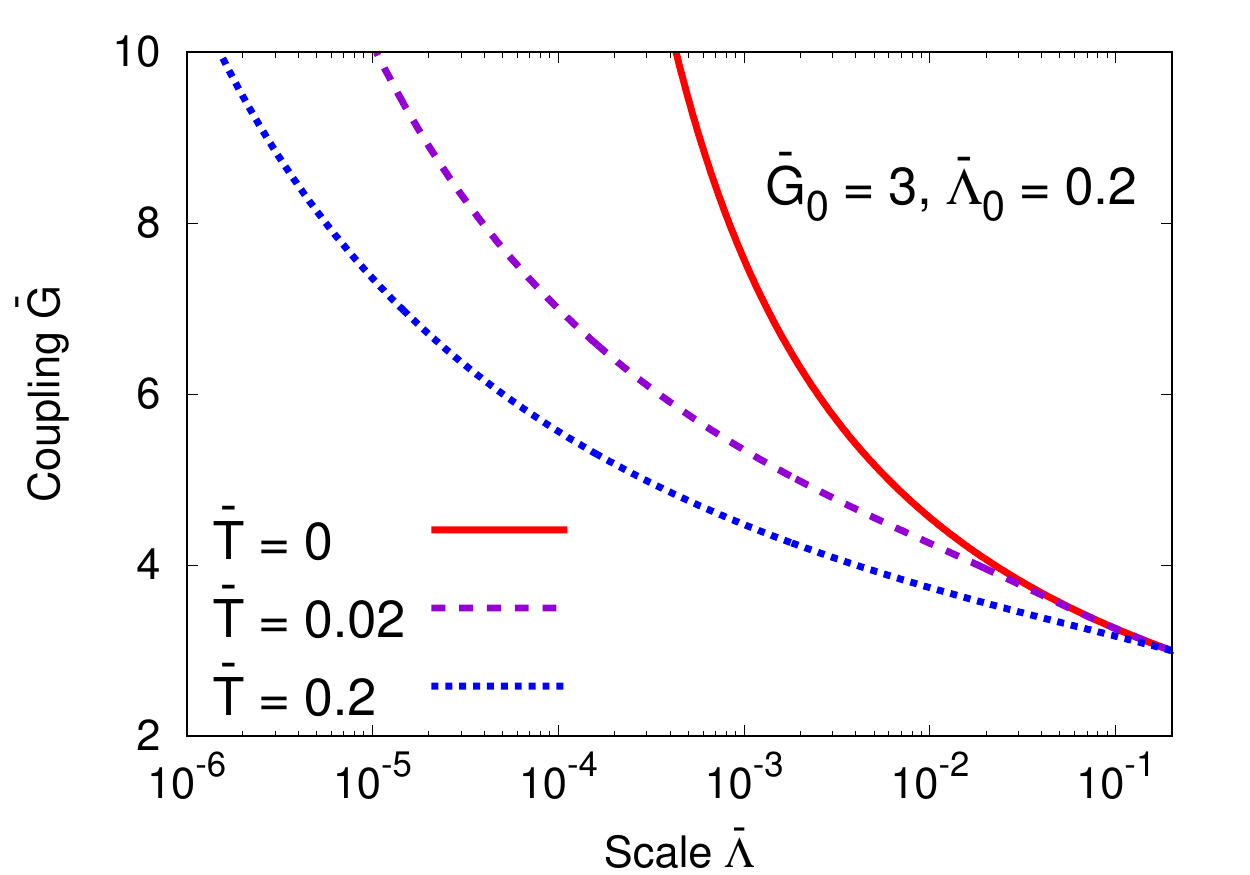}
\caption{The RG flow of the (dimensionless) coupling $\bar{G}$ with $N=3$ for $\bar{T}=0$ (red), $\bar{T}=0.02$ (purple) and $\bar{T}=0.2$ (blue). The initial values are $\bar{G}_0=3$ at $\bar{\Lambda}_0=0.2$.}
\label{fig:RGEPlot}
\end{figure}

The resulting RG flow of the dimensionless coupling $\bar{G}$ with $N=3$ is shown in Fig.~\ref{fig:RGEPlot}.
In this plot, the results with $\bar{T}=0$, $\bar{T}=0.02$, and $\bar{T}=0.2$ are shown.
As an example, the initial values are taken to be $\bar{G}_0 \equiv \bar{G}(\Lambda_0)= 3$ at the initial high-energy scale $\Lambda_0=0.2$.
The results clearly show the logarithmic divergences at lower-energy scales and the emergence of the Landau poles at the energy scale $\bar{\Lambda}=\bar{\Lambda}_{K}$ lower than $\bar{\Lambda}_0$ (or temperature), implying the appearance of the Kondo effect.
We call $\bar{\Lambda}_{K}$ the Kondo scale.
This behavior is easily understood by the fact that the right-hand side of Eq.~(\ref{RGEMu52}) is always negative.
It is important to note that the Kondo scale is generated dynamically through the quantum processes accompanying the non-Abelian interaction.\footnote{If there is no non-Abelian interaction (or the generator $t^{a}$) in the Lagrangian~\eqref{PLagrangian}, all the logarithmic divergences from ${\cal M}^{(1)}$ in Eqs.~\eqref{MOneComp} are canceled, and hence the Kondo scale disappears.}
The existence of the Kondo scale is more clearly confirmed in the case of zero temperature ($\bar{T}=0$).
In fact, from Eq.~\eqref{RGEMu52_0}, we obtain the analytic form of the solution as
\begin{eqnarray}
   \bar{G}(\bar{\Lambda}) = \frac{\bar{G}_{0}}{1+\frac{N\bar{G}_{0}}{8\pi^{2}} \ln \frac{\bar{\Lambda}}{\bar{\Lambda}_{0}}},
\end{eqnarray}
leading to
\begin{eqnarray}
   \bar{\Lambda}_{K}=\bar{\Lambda}_{0}e^{-\frac{8\pi^{2}}{N\bar{G}_{0}}} \ll \bar{\Lambda}_{0}.
\end{eqnarray}
The last inequality indicates that the Kondo scale is the low-energy scale, so that it is exponentially smaller than the high-energy scale $\bar{\Lambda}_{0}$.
At finite temperature, we notice that, as the temperature becomes higher, the value of $\bar{\Lambda}_{K}$ becomes smaller.
Thus, this behavior implies the suppression of the Kondo effect by finite temperature effects.

\section{Mean-field approach} \label{Sec_3}
At the low-energy scale below the Kondo scale, we need to describe the Kondo effect in a nonperturbative way.
For this purpose, we adopt a mean-field approach describing a mixing between a light relativistic fermion and a heavy fermion based on the treatment in Refs.~\cite{Yasui:2016svc,Yasui:2017izi}.

\subsection{Mean-field Lagrangian}
For the light relativistic fermions, we use the one-flavor light-fermion field $\psi$ with a chemical potential $\mu$ and a chiral chemical potential $\mu_5$.\footnote{The one flavor is a simplified setup, but
we can easily extend our formalism to multi-flavor fermions, $\psi \equiv(\psi_1^t, \psi_2^t, \cdots, \psi_{N_f}^t)$~\cite{Yasui:2016svc,Yasui:2017izi}.}
For the heavy fermions, we use a redefined field based on the so-called heavy-quark effective theory~\cite{Eichten:1989zv,Georgi:1990um} (see Refs.~\cite{Neubert:1993mb,Manohar:2000dt} for reviews): $\Psi_v \equiv \frac{1}{2} (1 + v^\mu \gamma_\mu) e^{i{M}_{Q} v \cdot x} \Psi$ where ${M}_{Q}$ and $v^\mu =(1,\vec{0})$ are the mass and four-velocity of the heavy fermion at rest (the rest frame), respectively.
After this redefinition, only the positive-energy component of the original Dirac spinor of the heavy-fermion field survives by the projection operator $ \frac{1}{2} (1 + \gamma_0)$.
The original mass ${M}_{Q}$ is subtracted by the factor $e^{i{M}_{Q} v \cdot x}$.

As a result, the effective Lagrangian is given by
\begin{eqnarray}
{\cal L} 
&=&\bar{\psi} ( i \partial\hspace{-0.55em}/ + \mu \,  \gamma_{0} + \mu_5  \gamma_{0} \gamma_5) \psi 
 + \bar{\Psi}_v i v^\mu \partial_\mu \Psi_v \nonumber \\
&& + \tilde{G} \Bigl[|\bar{\psi}_R \Psi_v|^2 + |\bar{\psi}_L \Psi_v|^2 + |\bar{\psi}_R \vec{\gamma} \Psi_v|^2 + |\bar{\psi}_L \vec{\gamma} \Psi_v|^2 \Bigr] \nonumber \\
&& - \lambda (\bar{\Psi}_v \Psi_v - n_Q ) , \label{eq:L2}
\end{eqnarray}
where the $SU(N)$ non-Abelian interaction term is a four-point vertex, and $\tilde{G}$ is the coupling constant in the interaction between a light fermion and a heavy particle.\footnote{Note that the four-point interaction in Eq.~\eqref{eq:L2} can be obtained by the Fiertz transformation from Eq.~\eqref{PLagrangian}.
See {\it e.g.}, Refs.~\cite{Yasui:2016svc,Yasui:2017izi}.
Using the projection operators for the chirality of the light fermions, $\psi_R = \frac{1+\gamma_5}{2} \psi$ and $\psi_L = \frac{1-\gamma_5}{2} \psi$, we can easily check the chiral symmetry for the four-point interaction terms:
\begin{eqnarray}
&& |\bar{\psi} \Psi_v|^2 + |\bar{\psi} i\gamma_5 \Psi_v|^2 + |\bar{\psi} \vec{\gamma} \Psi_v|^2 + |\bar{\psi} \gamma_5 \vec{\gamma} \Psi_v|^2 \nonumber \\
&=& 2\Bigl[|\bar{\psi}_R \Psi_v|^2 + |\bar{\psi}_L \Psi_v|^2 + |\bar{\psi}_R \vec{\gamma} \Psi_v|^2 + |\bar{\psi}_L \vec{\gamma} \Psi_v|^2 \Bigr]. \nonumber
\end{eqnarray}
}
$\lambda$ and $n_Q$ are the Lagrange multiplier and heavy-particle density, respectively, for the constraint condition $\bar{\Psi}_v \Psi_v = n_Q$~\cite{Yasui:2016svc,Yasui:2017izi}.\footnote{Notice that $\bar{\Psi}_{v}=\Psi_{v}^{\dag}$ in the rest frame.}
Thus, the number density of heavy particles are controlled by the Lagrange multiplier ($\lambda$) independent of light fermions, so that the heavy particles need not to satisfy chemical equilibrium conditions because they are impurities. Therefore, one needs not to regard $\lambda$ as a chemical potential of heavy particles.\footnote{The Kondo effect for a {\it single} heavy particle within the same mean-field ansatz is formalized in Ref.~\cite{Yasui:2016yet}.} The value of $n_Q$ is determined by solving a stationary condition of the thermodynamic potential: $\partial\Omega/\partial\lambda=0$ ($\Omega$ will be provided in Eq.~(\ref{omega}) explicitly). Here, we notice that choosing $\lambda=0$ does not necessarily impose $n_Q=0$.

As a mean-field approximation, we assume the following form of the condensate, which is the so-called {\it Kondo condensate}~\cite{Yasui:2016svc,Yasui:2017izi}:
\begin{eqnarray}
&& \tilde{G} \langle \bar{\psi}_R \Psi_v \rangle = \Delta_R, \ \ \ \tilde{G} \langle \bar{\psi}_L \Psi_v \rangle = \Delta_L, \label{eq:delta_S} \\
&& \tilde{G} \langle \bar{\psi}_R \vec{\gamma} \Psi_v  \rangle = \Delta_R \, \hat{p}, \ \ \ \tilde{G} \langle \bar{\psi}_L \vec{\gamma} \Psi_v  \rangle = \Delta_L \, \hat{p}, \label{eq:delta_V}
\end{eqnarray}
where $\hat{p} \equiv \vec{p}/p$ ($p \equiv |\vec{p}|$) is the unit vector for the three-dimensional momentum $\vec{p}$.\footnote{The momentum dependence in Eq.~\eqref{eq:delta_V} is called the {\it hedgehog} solution.
We assumed the scalar and hedgehog condensate have the same value of $\Delta_{R(L)}$.}
The angle brackets $\langle \mathcal{O} \rangle$ denote the vacuum expectation value for an operator $\mathcal{O}$.
Note that $\Delta_{R(L)}$ is a complex number, which indicates the mixing between the light fermion and the heavy particle.
Thus, $|\Delta_{R(L)}|$ gives the absolute value of the Kondo condensate.
From Eq.~\eqref{eq:L2}, as a result, the mean-field Lagrangian is written as 
\begin{eqnarray}
{\cal L}_{\mathrm{MF}} = \bar{\phi} \,  {\cal G}(p_{0},\vec{p}\,)^{-1} \phi - \frac{2 |\Delta_R|^{2}}{\tilde{G}} - \frac{2 |\Delta_L|^{2}}{\tilde{G}} + \lambda n_Q,
\label{eq:MF}
\end{eqnarray}
where $\phi \equiv (\psi^t, (\Psi_v^{\mathrm{pos}})^t)$ contains the six components with the Dirac four-spinor of the light-fermion field $\psi$ and the positive-energy projected components (two-spinor) of the heavy-particle field $\Psi_v^{t} \equiv ((\Psi_v^{\mathrm{pos}})^t,0)$.
The factor $2$ in front of $|\Delta_{R(L)}|^2$ comes from the 
ansatz (\ref{eq:delta_S}) and (\ref{eq:delta_V}).
The inverse propagator of $\phi$ is given by
\begin{widetext}
\begin{eqnarray}
 {\cal G}(p_{0},\vec{p}\,)^{-1}
\equiv
 \left(
\begin{array}{ccc}
p_0 + \mu                        & -\vec{p}\cdot \vec{\sigma} + \mu_5 & \frac{\Delta^\ast_R}{2}  (1+\hat{p}\cdot \vec{\sigma}) +\frac{\Delta^\ast_L }{2} (1-\hat{p}\cdot \vec{\sigma})  \\
\vec{p}\cdot \vec{\sigma} -\mu_5 & -(p_0 + \mu)                       & -\frac{\Delta^\ast_R}{2}  (1+\hat{p}\cdot \vec{\sigma}) +\frac{\Delta^\ast_L}{2}  (1-\hat{p}\cdot \vec{\sigma}) \\
\frac{\Delta_R}{2} (1+\hat{p}\cdot \vec{\sigma}) +\frac{\Delta_L}{2}  (1-\hat{p}\cdot \vec{\sigma})                            & \frac{\Delta_R}{2}  (1+\hat{p}\cdot \vec{\sigma}) -\frac{ \Delta_L}{2} (1-\hat{p}\cdot \vec{\sigma})    & p_0 - \lambda \\
\end{array}
\right), \nonumber\\
\label{eq:Ginverse_RL_Nf}
\end{eqnarray}
in the standard representation of the Dirac matrices.
\end{widetext}

Before closing this subsection, we comment the symmetry breaking pattern in the present analysis. Originally, the Lagrangian~(\ref{eq:L2}) possesses $U(1)_R\times U(1)_L$ chiral symmetry,  $SU(2)_{\rm HFS}$ heavy-fermion spin (HFS) symmetry, and $U(1)_h$ heavy-fermion number symmetry. Namely, the original global symmetry is $G=U(1)_R\times U(1)_L\times SU(2)_{\rm HFS}\times U(1)_h $. After the Kondo condensate in Eqs.~(\ref{eq:delta_S}) and~(\ref{eq:delta_V}) dominates the ground state, the symmetry will be broken to be $H= U(1)_{R+L+h}$ if $\Delta_R \neq \Delta_L$, where the $U(1)_{R+L+h}$ symmetry is associated with a conservation of the sum of the light-fermion number and the heavy-fermion number. As a particular case, if the Kondo condensate satisfies $\Delta_R=\Delta_L$, then the remaining global symmetry is $H = U(1)_{R+L+h} \times U(1)_{R-L+ {\rm HFS}_h} $, where the $U(1)_{R-L+ {\rm HFS}_h}$ stands for the so-called {\it chiral-HFS locked ($\chi$HFSL) symmetry} argued in Refs.~\cite{Yasui:2016svc,Yasui:2017izi}.

\subsection{Dispersion relations}
By solving $\det[{\cal G}(p_0,\vec{p})^{-1}]=0$, we obtain the six energy-momentum dispersion relations
\begin{eqnarray}
E_{R\pm}(p) &\equiv& \frac{1}{2} \Bigl( p + \lambda -\mu_{R} \pm \sqrt{\bigl(p-\lambda-\mu_{R} \bigr)^2 + 8 |\Delta_R|^2 }\Bigr),  \nonumber \\
\label{eq:mode_ER} \\
E_{L\pm}(p) &\equiv&  \frac{1}{2} \Bigl( p + \lambda -\mu_{L} \pm \sqrt{\bigl(p-\lambda-\mu_{L} \bigr)^2 + 8 |\Delta_L|^2 }\Bigr), \nonumber \\
\label{eq:mode_EL} \\
\tilde{E}_R(p) &\equiv& - p - \mu_{R}, \label{eq:mode_ERt} \\
\tilde{E}_L(p) &\equiv& - p - \mu_{L}.  \label{eq:mode_ELt}
\end{eqnarray}
with $\mu_{R,L}\equiv\mu\pm\mu_{5}$.
The four modes, $E_{R\pm}$ and $E_{L\pm}$, are the mixing modes (quasiparticles) between the light fermion and the heavy particle, which are induced by the nonzero value of the Kondo condensate $\Delta_{R(L)}$.
On the other hand, $\tilde{E}_R$ and $\tilde{E}_L$ are the decoupling anti-particle modes.
The obtained dispersion realtions and the wave functions lead to the quasiparticle fermions,
but they preserve the topological properties for the original massless Dirac fermions,
where the Berry's curvature induces the monopoles in momentum space~\cite{Yasui:2017izi}.

\begin{figure}[tb]
    \begin{minipage}[t]{1.0\columnwidth}
        \begin{center}
            \includegraphics[clip, width=0.7\columnwidth]{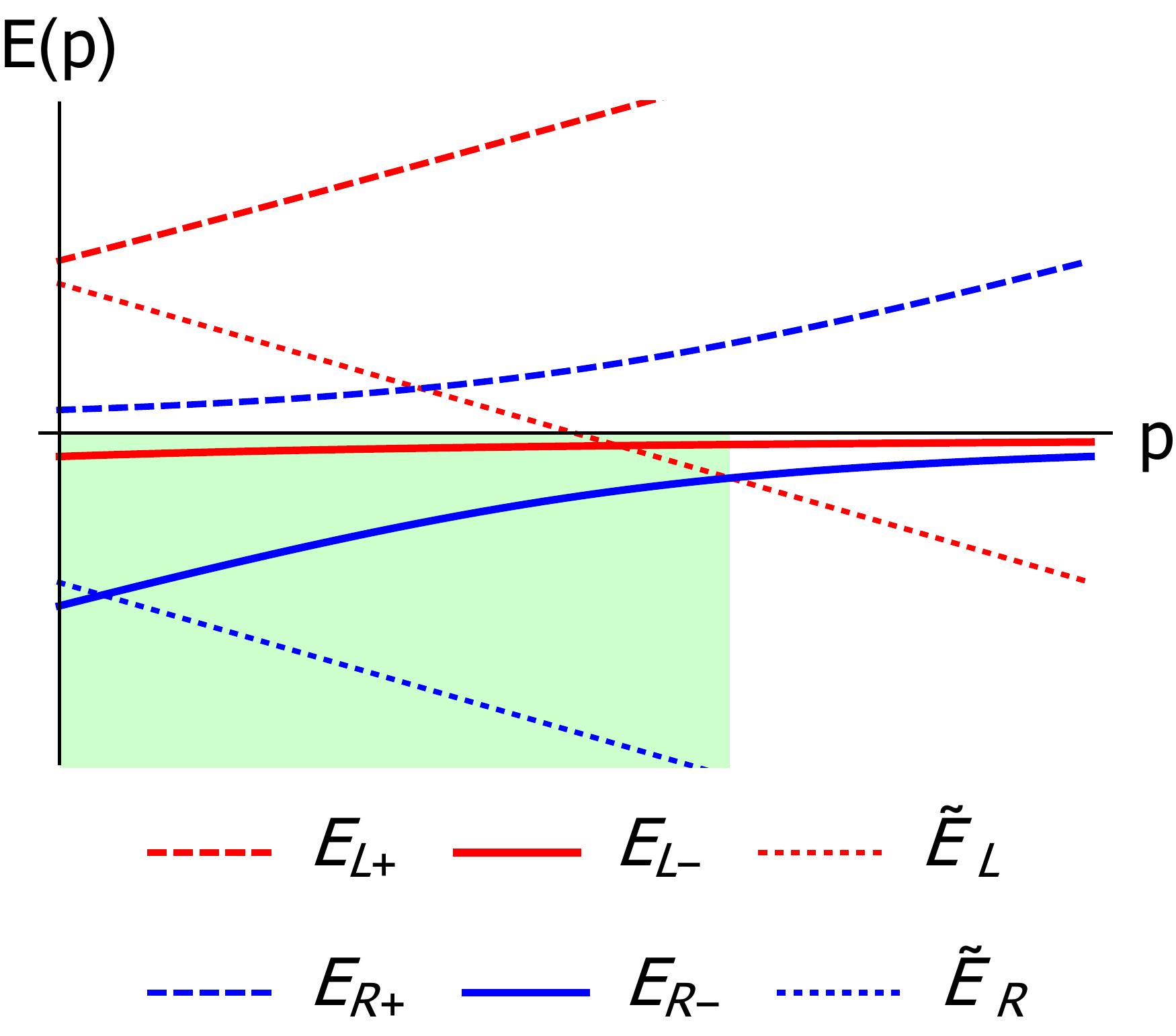}
        \end{center}
    \end{minipage}
    \caption{Dispersion relations of quasiparticles [$E_{R\pm}(p)$, $E_{L\pm}(p)$, $\tilde{E}_R(p)$, and $\tilde{E}_L(p)$] with Kondo condensate $\Delta_{R(L)}$ at finite $\mu_5$. The shadow area [$E(p)<0$] is the region where the (quasi)particles are occupied up to the cutoff momentum.
Here, we set $\Delta_R=\Delta_L$ as an example.}
\label{fig:disp}
\end{figure}

A schematic figure of these dispersion relations is shown in Fig.~\ref{fig:disp}.
Among them, the quasiparticles with $E_{R-}$ and $E_{L-}$ are essential for the Kondo effect
because the Kondo condensate is induced by the occupation of quasiparticles under $E(p)=0$.

\subsection{Thermodynamic potential}
From the modes in Eqs.~\eqref{eq:mode_ER}-\eqref{eq:mode_ELt},
the thermodynamic potential at finite temperature $T$ is obtained as
\begin{eqnarray}
\Omega(T,\mu,\mu_5,\lambda;\Delta_{R(L)})
 &=& N \int_0^{\Lambda_\mathrm{cut}} \hspace{-0.5em} 
 f(T,\mu,\mu_5,\lambda;p)
 \frac{p^2 dp}{2\pi^2} \nonumber \\
&& + \frac{2|\Delta_R|^2}{\tilde{G}} + \frac{2|\Delta_L|^2}{\tilde{G}}
 - \lambda n_Q,   \label{omega} 
\end{eqnarray}
where $\Lambda_\mathrm{cut}$ is an ultraviolet cutoff parameter of the momentum integral, and the integrand is
\begin{eqnarray}
&& f(T,\mu,\mu_5,\lambda;p) \nonumber\\
&=& - \frac{1}{2} \sum_{i= R,L} \left[ E_{i+}(p) + E_{i-}(p)+ \tilde{E}_{i}(p) \right] \nonumber\\
&& -\frac{1}{\beta}
\ln \biggl[
\prod_{i=R,L}
\left(1+e^{-\beta E_{i+}(p)}\Bigr) \Bigl(1+e^{-\beta E_{i-}(p)} \Bigr) \right. \nonumber\\
&& \left. \times \Bigl(1+e^{-\beta \tilde{E}_{i}(p)} \right)
\biggr].
\end{eqnarray}
From the minimization condition of Eq.~(\ref{omega}) or the gap equation ${\partial \Omega}/{\partial \Delta_R}={\partial \Omega}/{\partial \Delta_L}=0$, we can determine $\Delta_{R(L)}$ in a self-consistent way.
In this model setting, the free parameters are $\tilde{G}$ and $\Lambda_\mathrm{cut}$ (and $N$), and they can be tuned for a specific system, as it will be explained later.

\subsection{Numerical results} \label{Sec_3-4}
\begin{figure}[t!]
    \begin{minipage}[t]{1.0\columnwidth}
        \begin{center}
            \includegraphics[clip, width=1.0\columnwidth]{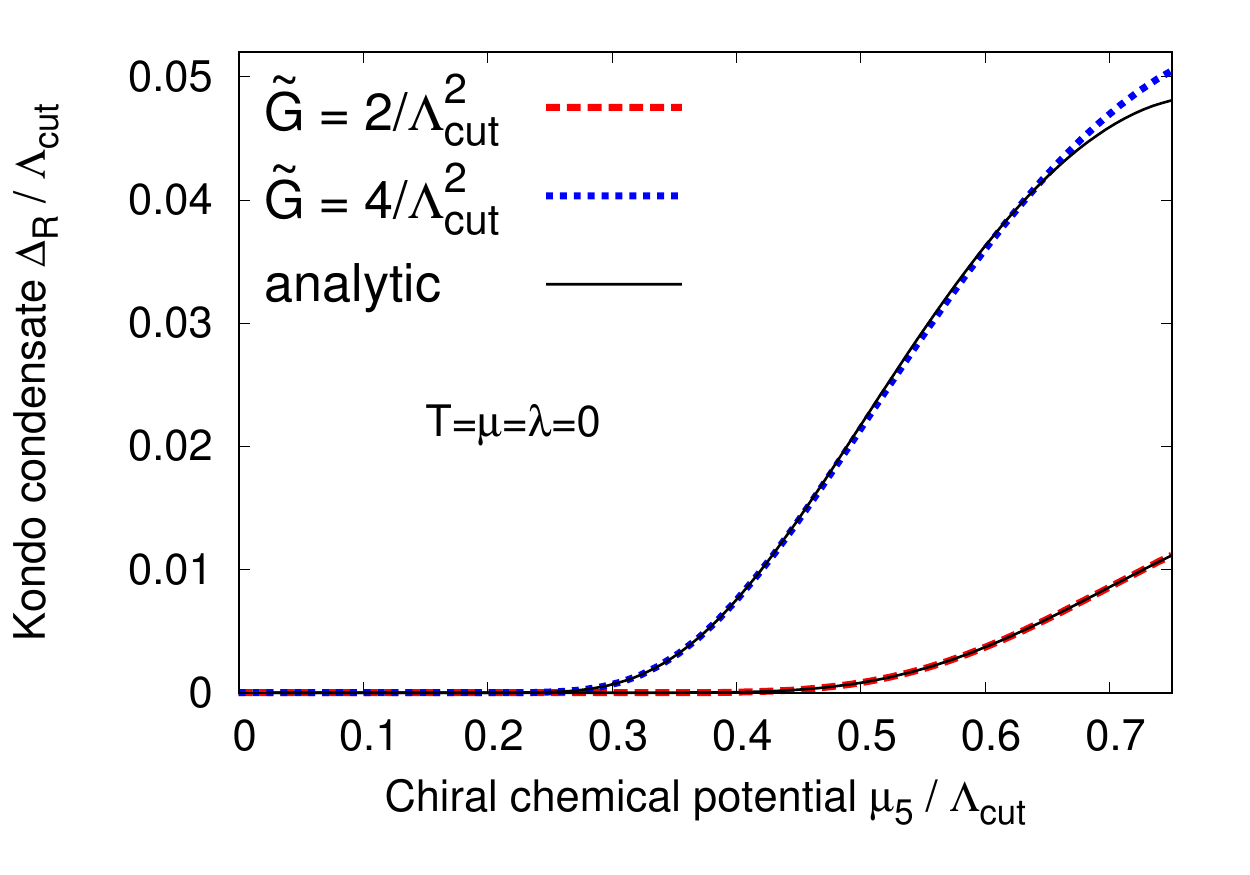}
        \end{center}
    \end{minipage}
    \caption{Kondo condensate $\Delta_R$ at finite $\mu_5>0$ and $T=\mu=\lambda=0$ using $\tilde{G}=2/\Lambda_\mathrm{cut}^2$ or $\tilde{G}=4/\Lambda_\mathrm{cut}^2$ and $N=3$.
Note that $\Delta_L \approx 0$ within these parameters.
{The black curves are the results from an analytic solution (\ref{DeltaRZeroT}).}}
\label{fig:mu5_delta}
\end{figure}

The Kondo condensate $\Delta_R$ as a function of $\mu_5>0$ is plotted in Fig.~\ref{fig:mu5_delta}.
Here we use, for example, $\tilde{G}=2/\Lambda^2_\mathrm{cut}$ and $4/\Lambda^2_\mathrm{cut}$ at $N=3$.
We find that $\Delta_R$ is enhanced as $\mu_5$ increases.
This behavior indicates that the (relativistic) Kondo effect is induced by finite $\mu_5$.
This is consistent with the result from the perturbative analysis in Sec.~\ref{Sec_2}.
We emphasize that the usual (nonrelativistic and relativistic) Kondo effects occur at finite $\mu$, but the Kondo effect at finite $\mu_5$ appears even when $\mu=0$. See Appendix~\ref{App:finite_mu} for the discussion at finite $\mu$.
This is a unique property of relativistic fermions composing matter including impurities.
Such Kondo effects can be realized in relativistic-fermion matter, {\it i.e.} Weyl/Dirac metal/semimetals and quark matter.

Within our parameters, we numerically find that, for $\mu_5>0$, the Kondo effect is dominated by the right-handed condensate $\Delta_R$, and the value of the left-handed condensate $\Delta_L$ is almost zero.
On the other hand, in the case of $\mu_5<0$, $\Delta_L$ dominates the Kondo effect.

For a typical parameter in the QCD Kondo effect, we apply the coupling constant, $\tilde{G} = G_c$, where $G_c \equiv 2 / \Lambda_\mathrm{cut}^2$ and $\Lambda_\mathrm{cut} = 0.65 \ \mathrm{GeV}$, and the number of the colors is $N=3$.
These parameters are the same as those used in the Nambu--Jona-Lasinio model with a four-point interaction between a light quark and a light antiquark~\cite{Klevansky:1992qe}.
When we use $G_c$, we find $\Delta_R=7.9 \, \mathrm{MeV}$ at $\mu_5 = 0.5 \, \mathrm{GeV}$.
If we use a stronger coupling constant, the Kondo effect is increasingly enhanced, as shown by the blue curve in Fig.~\ref{fig:mu5_delta}.
Note that, if we extrapolate the results to $0.75 \lesssim \mu_5/\Lambda_\mathrm{cut}$, then we find a sudden decrease of $\Delta_R$, but this behavior is an artifact from the cutoff $\Lambda_\mathrm{cut}$ in our model.

For a better understanding of the plot in Fig.~\ref{fig:mu5_delta}, here, we show the analytic expressions of $\mu_5$ dependence of $\Delta_R$ and $\Delta_L$. Under an assumption of $\Delta_R,\Delta_L\ll \mu_5,\Lambda_{\rm cut}$ with $T=\mu=\lambda=0$, the gap equation is solved analytically as
\begin{eqnarray}
\Delta_R &\approx& \alpha\sqrt{\frac{\mu_5(\Lambda_{\rm cut}-\mu_5)}{2}}{\rm exp}\left(-\frac{\pi^2}{N\mu_5^2\tilde{G}}\right)\ , \label{DeltaRZeroT} \\
\Delta_L &\approx& 0\ ,
\end{eqnarray}
with $\alpha = {\rm exp}\left[(\Lambda_{\rm cut}^2+2\Lambda_{\rm cut}\mu_5-6\mu_5^2)/(4\mu_5^2)\right]$, as shown in Refs.~\cite{Yasui:2016svc,Yasui:2017izi}. Thus, the value of $\Delta_R$ at $\mu_5/\Lambda_{\rm cut} \lesssim 0.3$ does not vanish but is simply suppressed exponentially. 
The analytic solution of $\mu_5$ dependence of $\Delta_R$ in Eq.~(\ref{DeltaRZeroT}) is shown by the black curve in Fig.~\ref{fig:mu5_delta}, which is in good agreement with the numerical result.

We comment the possible setup on lattice QCD simulations.
At finite $\mu$, the Monte-Carlo simulations suffer from the sign problem, so that it is difficult to measure the QCD Kondo effect (by finite $\mu$) by using lattice simulations.
On the other hand, at finite $\mu_5$, we can escape from the sign problem~\cite{Fukushima:2008xe,Yamamoto:2011gk,Yamamoto:2011ks,Braguta:2015zta,Braguta:2015owi,Astrakhantsev:2019wnp}, and the QCD Kondo effect (by finite $\mu_5$) will be observed.

\begin{figure}[t!]
    \begin{minipage}[t]{1.0\columnwidth}
        \begin{center}
            \includegraphics[clip, width=1.0\columnwidth]{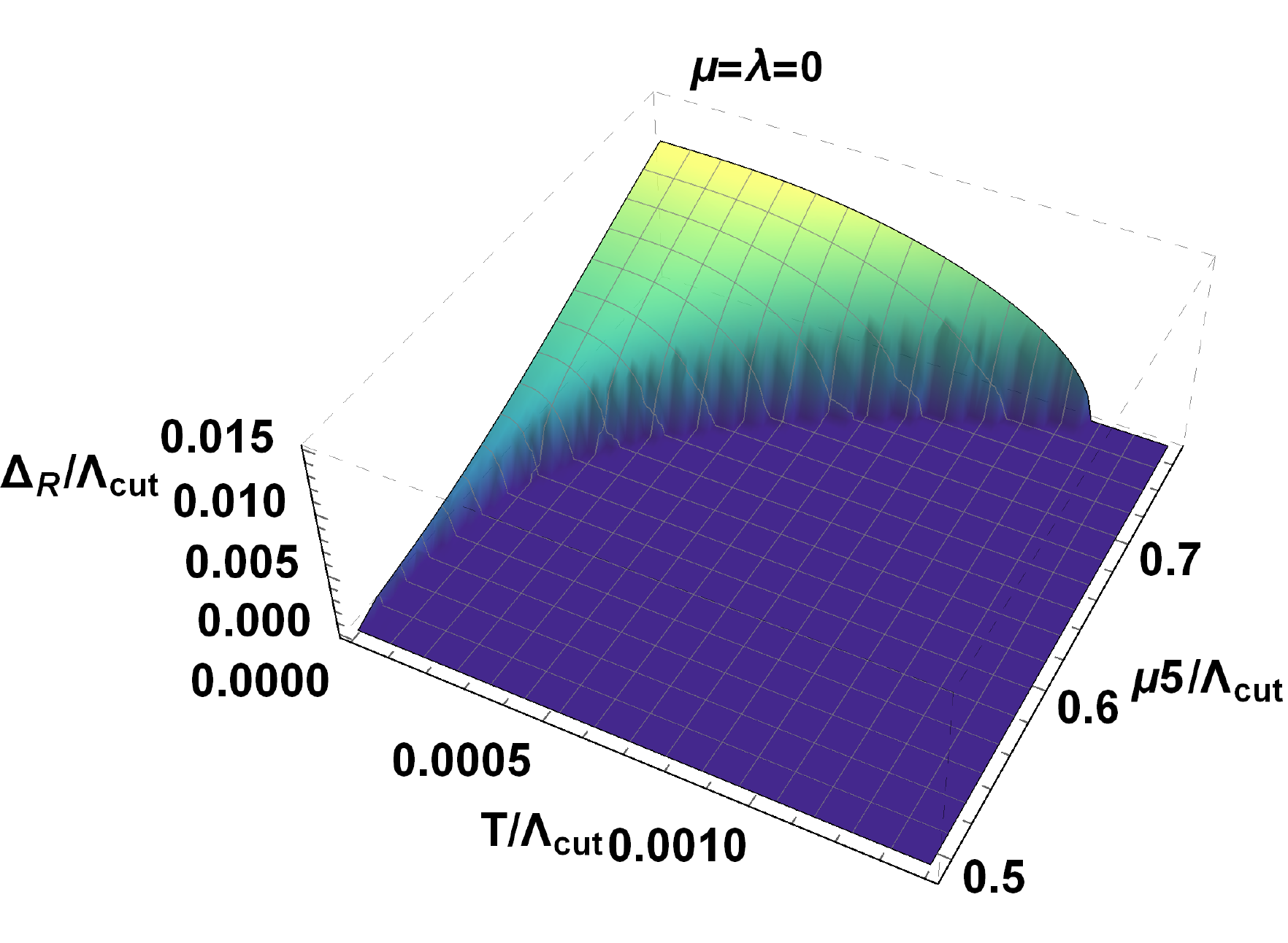}
        \end{center}
    \end{minipage}
    \caption{The phase diagram on the $T$-$\mu_5$ plane for the Kondo condensate $\Delta_R$ at $\mu=\lambda=0$. The parameters, $\tilde{G}=2/\Lambda_\mathrm{cut}^2$ and $N=3$, are used.
Note that $\Delta_L \approx 0$ within this parameter.}
\label{fig:T-mu5}
\end{figure}

Finally, we give a discussion on the temperature dependence of $\Delta_R$ at finite $\mu_5$.
In Fig.~\ref{fig:T-mu5}, we show $\Delta_R$ on the $T$-$\mu_5$ plane.
We observe that, when a finite $T$ is switched on, the value of $\Delta_R$ decreases: the Kondo effect is suppressed by finite-temperature effects, which is again consistent with the perturbative analysis in Sec.~\ref{Sec_2}.
The order of the phase transition at finite $T>0$ is of second order (see Appendix \ref{App:order} for examination based on a susceptibility).

\section{Conclusion and outlook} \label{Sec_4}
In this paper, we proposed the Kondo effect driven by a chirality imbalance (or chiral chemical potential $\mu_5$) from the point of view of the two theoretical approaches.
Using the perturbative approach, we found the infrared divergence of scattering amplitude as a signal of the Kondo effect.  
Using the mean-field approach, we found that the Kondo condensate is enhanced by finite $\mu_5$.
These are universal properties in relativistic-fermion matter with heavy impurities and a chirality imbalance, which can be attributed to the enhancement of the density of states at the Fermi surface.
Our findings generalize the analysis of the Kondo effect in Dirac or Weyl electron systems with an energy splitting among Dirac cones~\cite{Mitchell:2015}, involving various types of $SU(N)$ exchange interactions, such as spin, isospin, and color.
The interplay effect between the exchange interaction and particular spin-orbit coupling in crystalline electron systems, such as topological Dirac semimetal Cd${}_{3}$As${}_{2}$, is left for further analysis.

As a topics not covered in the present study, we comment that the response to magnetic and electric fields would be interesting.
For example, when $\mu_5$ is coupled to a magnetic field, an electric current can be induced, which is the so-called chiral magnetic effect~\cite{Kharzeev:2007jp,Fukushima:2008xe}.
The correlation between the chiral transport phenomena and the Kondo effects will be worth to be studied.
See for example the discussion of the transport coefficients in the Kondo effect in relativistic-fermion gas~\cite{Yasui:2017bey}.

In the context of QCD, lattice simulations at finite $\mu_5$ evade from the sign problem~\cite{Fukushima:2008xe,Yamamoto:2011gk,Yamamoto:2011ks,Braguta:2015zta,Braguta:2015owi,Astrakhantsev:2019wnp}, so that we can numerically measure the QCD Kondo effects in a fully nonperturbative way.
The ground state of QCD in the low-temperature and/or low-chemical potential region is the chiral-symmetry breaking phase characterized by the chiral condensate, and the ground state in the high-chemical potential region is expected to be the color superconducting phase characterized by diquark condensate.
These condensates could exclude the Kondo condensate~\cite{Kanazawa:2016ihl,Suzuki:2017gde} or might induce a ``coexistence" phase with two order parameters~\cite{Suzuki:2017gde}.
The topological properties of the QCD Kondo effect is also an interesting issue~\cite{Yasui:2017izi}.
However, the conclusion from the effective models depends on the coupling constants of the interactions, and in the future it should be checked based on QCD.

In particular, the properties of chiral condensates at finite $\mu_5$ have been studied from chiral effective models~\cite{Fukushima:2008xe,Fukushima:2010fe,Chernodub:2011fr,Ruggieri:2011xc,Gatto:2011wc,Andrianov:2012dj,Andrianov:2013dta,Buividovich:2014dha,Yu:2015hym,Braguta:2016aov,Frasca:2016rsi,Ruggieri:2016ejz,Farias:2016let,Cui:2016zqp,Lu:2016uwy,Ruggieri:2016xww,Pan:2016ecs,Khunjua:2018sro,Khunjua:2019lbv,Braguta:2019pxt,Das:2019crc,Yang:2019lyn}, Schwinger-Dyson equations~\cite{Wang:2015tia,Xu:2015vna}, and lattice QCD simulations~\cite{Braguta:2015zta,Braguta:2015owi,Astrakhantsev:2019wnp}.
One of the characteristic properties is the catalysis effect of the chiral symmetry breaking by finite $\mu_5$.
Therefore, in matter with a chirality imbalance and impurities, the two catalysis effects of the chiral symmetry breaking and Kondo effect could be correlated.

If we attempt to experimentally observe the Kondo effect in environments with a chirality imbalance, the finite-temperature effect will be practically important.
In particular, high-energy HICs produce high-temperature medium, and it could suppress the Kondo effect.
The melting temperature of Kondo effect estimated in this paper will be useful for future study.

In addition, in two- (or multi-) component fermion systems, the situation including an imbalance between the chemical potentials of different fermions would be also important.
In QCD, the isospin chemical potential $\mu_I$, an imbalance between up- and down- quark chemical potentials, is realized in neutron-rich nuclei and neutron stars, and lattice QCD simulations are also applicable~\cite{Kogut:2002tm,Kogut:2002zg,Kogut:2004zg,Brandt:2017oyy}.
For a similar external parameter to $\mu_5$, the effects from the chiral isospin chemical potential $\mu_{I5}$ could be also interesting~\cite{Ebert:2016hkd,Khunjua:2017khh,Khunjua:2017mkc,Khunjua:2018sro,Khunjua:2018jmn,Khunjua:2019ini,Khunjua:2019lbv}.

\section*{Acknowledgments}
Y.~A. is supported by the Leading Initiative for Excellent Young Researchers (LEADER).
This work is supported by National Natural Science Foundation of China (NSFC) Grant 20201191997 (D.~S.), and by Japan Society for the Promotion of Science (JSPS) Grant-in-Aid for Scientific Research (KAKENHI Grants No.~JP17K14316 (Y.~A.), No.~JP17K14277 (K.~S.) and No.~JP17K05435 (S.~Y.)), and by the Ministry of Education, Culture, Sports, Science and Technology (MEXT)-Supported Program for the Strategic Research Foundation at Private Universities ``Topological Science" (Grant No. S1511006) (S.~Y.).

\appendix
\section{Matsubara summation in Eqs.~(\ref{POneLoop1}) and~(\ref{POneLoop2}).}
\label{sec:MatsubaraSum}

In this appendix, we show a detailed calculation of Matsubara summation in the one-loop amplitudes in Eqs.~(\ref{POneLoop1}) and~(\ref{POneLoop2}). 

Within the imaginary-time formalism, Eqs.~(\ref{POneLoop1}) and~(\ref{POneLoop2}) are rewritten to
\begin{widetext}
\begin{eqnarray}
{\cal M}^{(1a)} &=& -G^2 \sum_{\epsilon_5=\pm}T\sum_{n}\int\frac{d^3k}{(2\pi)^3} \bar{u}(p_f) t^a\gamma^\mu P_{\epsilon_5}\tilde{\Delta}_l\big(i(\omega_{n}-i\epsilon_5\mu_5)\big)t^b\gamma^\nu u(p_i) \bar{U}(q_f)t^a\gamma_\mu \tilde{\Delta}_h(i\omega_{q_i}-i\omega_{n}+i\omega_{p_i})t^b\gamma_\nu U(q_i) \ , \nonumber\\\label{M1Appendix1}
\end{eqnarray}
and
\begin{eqnarray}
{\cal M}^{(1b)} &=& -G^2 \sum_{\epsilon_5=\pm}T\sum_{n}\int\frac{d^3k}{(2\pi)^3} \bar{u}(p_f) t^a\gamma^\mu P_{\epsilon_5}\tilde{\Delta}_l\big(i(\omega_{n}-i\epsilon_5\mu_5)\big)t^b\gamma^\nu u(p_i) \bar{U}(q_f)t^b \gamma_\nu\tilde{\Delta}_h(i\omega_{q_i}+i\omega_{n}-i\omega_{p_f})t^a\gamma_\mu U(q_i)  \ , \nonumber\\\label{M1Appendix2}
\end{eqnarray}
\end{widetext}
respectively, where the Matsubara Green's functions for the light and heavy fermions are given by
\begin{eqnarray}
\tilde{\Delta}_l\big(i(\omega_{n}-i\epsilon_5\mu_5)\big) = -\frac{i(-\omega_{n}+i\epsilon_5\mu_5)\gamma_0+\vec{k}\cdot\vec{\gamma}}{(\omega_{n}-i\epsilon_5\mu_5)^2+|\vec{k}|^2}\ ,   \nonumber\\ \label{DeltaLDefApp}
\end{eqnarray}
and
\begin{eqnarray}
\tilde{\Delta}_h(i\omega_{n}) = -\frac{-i\omega_{n}\gamma_0+\vec{k}\cdot\vec{\gamma}-M_Q}{\omega_{n}^2+|\vec{k}|^2 + M_Q^2}\ , \label{DeltaHDefApp}
\end{eqnarray}
with the Matsubara frequency $\omega_{n} = (2n+1)\pi T$ ($n=0,\pm1,\pm2,\cdots$).
Therefore, apart from the spinor and $SU(N)$ non-Abelian algebras, we need to calculate 
\begin{eqnarray}
{\cal I}_1 &\equiv&T\sum_{n}\int\frac{d^3k}{(2\pi)^3}\tilde{\Delta}_l\big(i(\omega_{n}-i\epsilon_5\mu_5)\big) \nonumber\\
&& \otimes\tilde{\Delta}_h(i\omega_{q_i}-i\omega_{n}+i\omega_{p_i}) \ ,\label{I1Def}
\end{eqnarray}
and
\begin{eqnarray}
{\cal I}_2 &\equiv&T\sum_{n}\int\frac{d^3k}{(2\pi)^3}\tilde{\Delta}_l\big(i(\omega_{n}-i\epsilon_5\mu_5)\big) \nonumber\\
&& \otimes\tilde{\Delta}_h(i\omega_{q_i}+i\omega_{n}-i\omega_{p_f})\ , \label{I2Def}
\end{eqnarray}
for the evaluation of ${\cal M}^{(1a)}$ and ${\cal M}^{(1b)}$.

First, let us demonstrate a detailed calculation of ${\cal I}_1$.
The three-momentum integral in Eq.~(\ref{I1Def}) is performed by the conventional procedure as in the vacuum, such that we show only the zeroth components of the momentum or coordinate space explicitly below.
The inverse Fourier transformations of the Matsubara Green's functions $\tilde{\Delta}_l\big(i(\omega_{n}-i\epsilon_5\mu_5)\big)$ and $\tilde{\Delta}_h(i\omega_{n})$ given in Eqs.~(\ref{DeltaLDefApp}) and~(\ref{DeltaHDefApp}) can be defined by
\begin{eqnarray}
\tilde{\Delta}_l\big(i(\omega_{n}-i\epsilon_5\mu_5)\big) &=& \int_0^\beta d\tau \tilde{\Delta}_l(\tau){e}^{i(\omega_{n}-i\epsilon_5\mu_5)\tau}, \nonumber\\
\tilde{\Delta}_h(i\omega_n) &=& \int_0^\beta d\tau \tilde{\Delta}_h(\tau){e}^{i\omega_{n}\tau}\ .
\end{eqnarray}
Then, by making use of the Poisson summation formula
\begin{eqnarray}
\sum_n\delta(x+\beta n) &=& T\sum_{n=-\infty}^\infty {e}^{i\frac{2\pi n}{\beta}x} \nonumber\\
&=& T\sum_{n=-\infty}^\infty {e}^{i\omega_{n} x}{e}^{-i\frac{\pi x}{\beta} }\ ,
\end{eqnarray}
we get the following equation:
\begin{eqnarray}
{\cal J}_1 &\equiv&T\sum_{n} \tilde{\Delta}_l(i(\omega_{n}-i\epsilon_5\mu_5)) \otimes\tilde{\Delta}_h(i\omega_{q_i}-i\omega_{n}+i\omega_{p_i}) \nonumber\\
&=& \int_0^\beta d\tau \tilde{\Delta}_l(\tau) \otimes\tilde{\Delta}_h(\tau){e}^{\epsilon_5\mu_5\tau+i(\omega_{q_i}+\omega_{p_i})\tau} \ . \label{J1}
\end{eqnarray}
The Matsubara Green's function $\tilde{\Delta}_{l(h)}(\tau)$ is defined by an analytic continuation of the greater Green's function
\begin{eqnarray}
S_{l}^>(t) &=& \langle \psi(t)\bar{\psi}(0)\rangle_\beta \ , \nonumber\\
S_h^>(t) &=& \langle\Psi(t)\bar{\Psi}(0) \rangle_\beta\ ,
\end{eqnarray}
as
\begin{eqnarray}
\tilde{\Delta}_{l(h)}(\tau) = S_{l(h)}^>(-i\tau) \ .
\end{eqnarray}
Here, we remind that the Fourier transformation of the greater Green's function $S_{l(h)}^>(t)$ can be expressed as~\cite{Bellac:2011kqa}
\begin{eqnarray}
S_{l}^>(t) &=& \int\frac{dk_0}{2\pi} \big(1-\tilde{f}_{\beta}(k_0- \epsilon_5 \mu_5 )\big) \tilde{\rho}_l(k_0) e^{-ik_0t} \ , \nonumber\\
S_{h}^>(t) &=& \int\frac{dk_0}{2\pi} \big(1-\tilde{f}_{\beta}(k_0)\big) \tilde{\rho}_h (k_0) e^{-ik_0t} \ ,
\end{eqnarray}
in which $\tilde{f}_{\beta}(k_0)$ is the Fermi distribution function,
$\tilde{f}_{\beta}(k_0)=1/(e^{\beta k_0}+1)$
($\beta=1/T$), and $\tilde{\rho}_{l(h)}$ is the spectral function 
\begin{eqnarray}
\tilde{\rho}_l(k_0) &=& 2\pi \epsilon(k_0)\Slash{k}\delta(k^2)\ , \nonumber\\
\tilde{\rho}_h(k_0) &=& 2\pi \epsilon(k_0)(\Slash{k}+M_Q)\delta(k^2-M_Q^2) \ .
\end{eqnarray}
Then, we find that Eq.~(\ref{J1}) can be rewritten to
\begin{widetext}
\begin{eqnarray}
{\cal J}_1 &=&  \int_0^\beta d\tau S_l^>(-i\tau)\otimes S^>_h(-i\tau) e^{(\epsilon_5\mu_5+i\omega_{q_i}+i\omega_{p_i})\tau}  \nonumber\\
&=& \int\frac{dk_0}{2\pi}\frac{dk_0'}{2\pi}\frac{1-\tilde{f}_{\beta}(k_0-\epsilon_5\mu_5)-\tilde{f}_{\beta}(k_0')}{k_0+k_0'-i\omega_{q_i}-i\omega_{p_i}-\epsilon_5\mu_5}\tilde{\rho}_l(k_0) \otimes \tilde{\rho}_h(k_0')\nonumber\\
&=& -\frac{1}{4|\vec{k}|E_{k'}}\Bigg[\frac{\big(1-\tilde{f}_{\beta}(|\vec{k}|-\epsilon_5\mu_5)-\tilde{f}_{\beta}(E_{k'})\big)F(|\vec{k}|;E_{k'})}{i\omega_{q_i}+i\omega_{p_i}+\epsilon_5\mu_5-|\vec{k}| - E_{k'}} + \frac{\big(\tilde{f}_{\beta}(|\vec{k}|-\epsilon_5\mu_5)-\tilde{f}_{\beta}(E_{k'})\big)F(|\vec{k}|;-E_{k'})}{i\omega_{q_i}+i\omega_{p_i}+\epsilon_5\mu_5-|\vec{k}|+E_{k'}} \nonumber\\
&& -\frac{\big(\tilde{f}_{\beta}(|\vec{k}|+\epsilon_5\mu_5)-\tilde{f}_{\beta}(E_{k'})\big) F(-|\vec{k}|,E_{k'})}{i\omega_{q_i}+i\omega_{p_i}+\epsilon_5\mu_5+|\vec{k}|-E_{k'}} - \frac{\big(1-\tilde{f}_{\beta}(|\vec{k}|+\epsilon_5\mu_5)-\tilde{f}_{\beta}(E_{k'})\big) F(-|\vec{k}|;-E_{k'})}{i\omega_{q_i}+i\omega_{p_i}+\epsilon_5\mu_5+|\vec{k}|+E_{k'}}\Bigg]\ , \label{I1Summed}
\end{eqnarray}
with $E_{k'}=\sqrt{|\vec{q}_i-\vec{k}+\vec{p}_i|^2+M_Q^2}$, where $F(k_0;k_0')$ is defined as
$F(k_0;k_0') = \Slash{k} \otimes (\Slash{k}'+M_Q)$.
By performing the analytic continuations of $i\omega_{q_i} \to q_i^0+i\epsilon$, $i\omega_{p_i} \to p_i^0+i\epsilon$, and replacing the energy of the external heavy fermion by its mass as $q_0 \to M_Q$, together with the $M_Q\to\infty$ limit, we find that Eq.~(\ref{I1Summed}) is reduced to
\begin{eqnarray}
{\cal J}_1 
\approx  -\frac{1}{2|\vec{k}|}\Bigg[\frac{1-\tilde{f}_{\beta}(|\vec{k}|-\epsilon_5\mu_5)}{p_i^0+\epsilon_5\mu_5-|\vec{k}| +i\epsilon}(|\vec{k}|\gamma^0-\vec{k}\cdot\vec{\gamma})\otimes \bar{\Lambda}_+  -\frac{\tilde{f}_{\beta}(|\vec{k}|+\epsilon_5\mu_5)}{p_i^0+\epsilon_5\mu_5+|\vec{k}|+i\epsilon} (-|\vec{k}|\gamma^0-\vec{k}\cdot\vec{\gamma})\otimes \bar{\Lambda}_+ \Bigg] \ ,\nonumber\\
\end{eqnarray}
with $\bar{\Lambda}_+ \equiv \lim_{M_Q\to\infty} (\Slash{q}+M_Q)/(2M_Q)=(1+\gamma_0)/2$.
Hence, by combining the three-momentum integral, finally we can evaluate ${\cal I}_1$ in Eq.~(\ref{I1Def}) as
\begin{eqnarray}
{\cal I}_ 1 \approx -\frac{1}{2}\int\frac{d^3k}{(2\pi)^3}\Bigg[\frac{1-\tilde{f}_{\beta}(|\vec{k}|-\epsilon_5\mu_5)}{(p^0_i+\epsilon_5\mu_5)-|\vec{k}|}+\frac{\tilde{f}_{\beta}(|\vec{k}|+\epsilon_5\mu_5)}{(p^0_i+\epsilon_5\mu_5)+|\vec{k}|} \Bigg]\gamma^0 \otimes \bar{\Lambda}_+\ .
\end{eqnarray}
Note that we are interested in only the real part of the amplitude, so that the imaginary parts have been omitted.
Therefore, ${\cal M}^{(1a)}$ in Eq.~(\ref{M1Appendix1}) becomes
\begin{eqnarray}
{\cal M}^{(1a)} &\approx& \frac{G^2}{2}\int\frac{d^3k}{(2\pi)^3}\Biggl[\frac{1-\tilde{f}_{\beta}(|\vec{k}|-\mu_5)}{p^0_i+\mu_5-|\vec{k}|}+\frac{\tilde{f}_{\beta}(|\vec{k}|+\mu_5)}{p^0_i+\mu_5+|\vec{k}|}\Biggr]\bar{u}(p_f)t^a\gamma^\mu P_+\gamma_0 t^b\gamma^\nu u(p_i) \bar{U}_+(q_f)t^a\gamma_\mu \bar{\Lambda}_+ t^b\gamma_\nu U_+(q_i)\nonumber\\
&&  + \frac{G^2}{2}\int\frac{d^3k}{(2\pi)^3}\Biggl[\frac{1-\tilde{f}_{\beta}(|\vec{k}|+\mu_5)}{p^0_i-\mu_5-|\vec{k}|} + \frac{\tilde{f}_{\beta}(|\vec{k}|-\mu_5)}{p^0_i-\mu_5+|\vec{k}|}\Biggr] \bar{u}(p_f)t^a\gamma^\mu P_-\gamma_0 t^b\gamma^\nu u(p_i) \bar{U}_+(q_f)t^a\gamma_\mu \bar{\Lambda}_+ t^b\gamma_\nu U_+(q_i) \ ,
\label{M1AppComp1}
\end{eqnarray}
by replacing $U(q_i) \to U_+(q_i)$ $\big(\bar{U}(q_f)\to \bar{U}_+(q_f)\big)$ together with the $M_Q\to\infty$ limit. 

In a similar manner, we can evaluate ${\cal I}_2$ in Eq.~(\ref{I2Def}) as
\begin{eqnarray}
{\cal I}_2 \approx- \frac{1}{2}\int\frac{d^3k}{(2\pi)^3}\Bigg[\frac{\tilde{f}_{\beta}(|\vec{k}|-\epsilon_5\mu_5)}{p^0_f+\epsilon_5\mu_5-|\vec{k}|} + \frac{1-\tilde{f}_{\beta}(|\vec{k}|+\epsilon_5\mu_5)}{p^0_f+\epsilon_5\mu_5+|\vec{k}|} \Bigg]\gamma^0 \otimes\bar{\Lambda}_+ \ ,
\end{eqnarray}
which yields that ${\cal M}^{(1b)}$ in Eq.~(\ref{M1Appendix2}) becomes
\begin{eqnarray}
{\cal M}^{(1b)} &\approx& \frac{G^2}{2}\int\frac{d^3k}{(2\pi)^3}\Biggl[\frac{\tilde{f}_{\beta}(|\vec{k}|-\mu_5)}{p^0_f+\mu_5-|\vec{k}|}+\frac{1-\tilde{f}_{\beta}(|\vec{k}|+\mu_5)}{p^0_f+\mu+|\vec{k}|}\Biggr] \bar{u}(p_f)t^a\gamma^\mu P_+\gamma_0t^b\gamma^\nu u(p_i) \bar{U}_+(q_f)t^b\gamma_\nu \bar{\Lambda}_+ t^a\gamma_\mu U_+(q_i)\nonumber\\ 
&& + \frac{G^2}{2}\int\frac{d^3k}{(2\pi)^3} \Biggl[\frac{\tilde{f}_{\beta}(|\vec{k}|+\mu_5)}{p^0_f-\mu_5-|\vec{k}|}+\frac{1-\tilde{f}_{\beta}(|\vec{k}|-\mu_5)}{p^0_f-\mu_5 + |\vec{k}|}\Biggr] \bar{u}(p_f)t^a\gamma^\mu P_-\gamma_0 t^b\gamma^\nu u(p_i) \bar{U}_+(q_f)t^b\gamma_\nu \bar{\Lambda}_+t^a\gamma_\mu U_+(q_i) \ ,  \nonumber\\
\label{M1AppComp2}
\end{eqnarray} 
in the same limit. The total one-loop amplitude is given by the sum of Eqs.~(\ref{M1AppComp1}) and~(\ref{M1AppComp2}): ${\cal M}^{(1)} = {\cal M}^{(1a)}+{\cal M}^{(1b)}$.

In the present study, we are interested only in the vicinity of the ``Fermi surface'' defined for the right-handed fermion with $\mu_5>0$. Namely, we assume that the initial- and finial-state light fermions satisfy the kinematics of $(p^0, |\vec{p}|) = (0,\mu_5)$ for the right-handed fermion while $(p^0, |\vec{p}|) = (2\mu_5,\mu_5)$ for the left-handed fermion ($p^\mu$  stands for $p^\mu_i$ and $p^\mu_f$ collectively), due to the Dirac equation. Thus, upon this assumption, ${\cal M}^{(1)}$ reads
\begin{eqnarray}
{\cal M}^{(1)} &\approx& \frac{G^2}{2}\int\frac{d^3k}{(2\pi)^3}\Biggl[\frac{1-\tilde{f}_{\beta}(|\vec{k}|-\mu_5)}{\mu_5-|\vec{k}|}+\frac{\tilde{f}_{\beta}(|\vec{k}|+\mu_5)}{\mu_5+|\vec{k}|}\Biggr]\bar{u}_R(p_f)t^a t^b  \gamma_0u_R(p_i) \bar{U}_+(q_f)t^at^b U_+(q_i)\nonumber\\
&&  + \frac{G^2}{2}\int\frac{d^3k}{(2\pi)^3}\Biggl[\frac{1-\tilde{f}_{\beta}(|\vec{k}|+\mu_5)}{\mu_5-|\vec{k}|} + \frac{\tilde{f}_{\beta}(|\vec{k}|-\mu_5)}{\mu_5+|\vec{k}|}\Biggr] \bar{u}_L(p_f)t^a t^b\gamma_0 u_L(p_i) \bar{U}_+(q_f)t^a t^b U_+(q_i) \nonumber\\ 
&&+\frac{G^2}{2}\int\frac{d^3k}{(2\pi)^3}\Biggl[\frac{\tilde{f}_{\beta}(|\vec{k}|-\mu_5)}{\mu_5-|\vec{k}|}+\frac{1-\tilde{f}_{\beta}(|\vec{k}|+\mu_5)}{\mu_5+|\vec{k}|}\Biggr] \bar{u}_R(p_f)t^at^b  \gamma_0u_R(p_i) \bar{U}_+(q_f)t^b t^a U_+(q_i)\nonumber\\ 
&& + \frac{G^2}{2}\int\frac{d^3k}{(2\pi)^3} \Biggl[\frac{\tilde{f}_{\beta}(|\vec{k}|+\mu_5)}{\mu_5-|\vec{k}|}+\frac{1-\tilde{f}_{\beta}(|\vec{k}|-\mu_5)}{\mu_5 + |\vec{k}|}\Biggr] \bar{u}_L(p_f)t^a t^b  \gamma_0 u_L(p_i) \bar{U}_+(q_f)t^bt^aU_+(q_i)\ .
\end{eqnarray}
When we choose $\mu_5>0$ and assume its value is large compared to the temperature $T$, but small enough so that $M_Q\to \infty$ limit is justified, the terms including $1/(|\vec{k}|+\mu_5)$ or $\tilde{f}_\beta(|\vec{k}|+\mu_5)$ in Eqs.~(\ref{M1AppComp1}) and~(\ref{M1AppComp2}) can be neglected. Hence, we find that ${\cal M}^{(1)}$ is reduced to
\begin{eqnarray}
{\cal M}^{(1)} 
&\approx& \frac{G^2}{2}\int\frac{d^3k}{(2\pi)^3}\frac{1-\tilde{f}_{\beta}(|\vec{k}|-\mu_5)}{\mu_5-|\vec{k}|}\bar{u}_R(p_f)t^at^b\gamma_0u_R(p_i) \bar{U}_+(q_f)t^at^bU_+(q_i)\nonumber\\
&& -\frac{G^2}{2}\int\frac{d^3k}{(2\pi)^3}\frac{\tilde{f}_{\beta}(|\vec{k}|-\mu_5)}{|\vec{k}|-\mu_5} \bar{u}_R(p_f)t^at^b\gamma_0u_R(p_i) \bar{U}_+(q_f)t^bt^aU_+(q_i)\nonumber\\ 
&& + \frac{G^2}{2}\int\frac{d^3k}{(2\pi)^3} \frac{1}{\mu_5-|\vec{k}|}\bar{u}_L(p_f)t^at^b\gamma_0u_L(p_i) \bar{U}_+(q_f)t^bt^aU_+(q_i) \ .  \label{M1AppComp}
\end{eqnarray} 
This expression clearly shows that the transition amplitude of the left-handed fermion is not affected by the Fermi surface, as naively anticipated. Then, by defining $E=|\vec{k}|-\mu_5$ for the first line in Eq.~(\ref{M1AppComp}) while $E=\mu_5-|\vec{k}|$ for the second line, we find
\begin{eqnarray}
{\cal M}^{(1)} & \approx& -\frac{G^2}{2}\rho_0\int_{-\mu_5}^\infty dE\frac{1-\tilde{f}_{\beta}(E)}{E}\bar{u}_R(p_f)t^at^b \gamma_0 u_R(p_i)\bar{U}_+(q_f)t^at^bU_+(q_i) \nonumber\\
&&+ \frac{G^2}{2}\rho_0\int_{-\mu_5}^{\infty} dE\frac{\tilde{f}_{\beta}(-E)}{E}\bar{u}_R(p_f)t^at^b \gamma_0 u_R(p_i)\bar{U}_+(q_f)t^bt^a U_+(q_i) \nonumber\\
&& + \frac{G^2}{2}\int\frac{d^3k}{(2\pi)^3} \frac{1}{\mu_5-|\vec{k}|}\bar{u}_L(p_f)t^at^b\gamma_0u_L(p_i) \bar{U}_+(q_f)t^bt^aU_+(q_i)  \ ,
\end{eqnarray}
\end{widetext}
where we have replaced the density of states by that on the Fermi surface,
$\rho_0 = \mu_5^2/(2\pi^2)$,
since we assumed a hierarchy of $M_Q (\to \infty) \gg \mu_5 \gg T$.

By using a relation $\tilde{f}_{\beta}(-E)=1-\tilde{f}_{\beta}(E)$ and the identities
\begin{eqnarray}
(t^at^b)_{kl}(t^at^b)_{ij} &=& \frac{N^2-1}{4N^2}\delta_{kl}\delta_{ij}-\frac{1}{N}(t^a)_{kl}(t^a)_{ij}\ , \nonumber\\
(t^at^b)_{kl}(t^bt^a)_{ij} &=& \frac{N^2-1}{4N^2}\delta_{kl}\delta_{ij}-\frac{2-N^2}{2N}(t^a)_{kl}(t^a)_{ij} \ , \nonumber\\
\end{eqnarray}
finally we arrive at
\begin{eqnarray}
{\cal M}^{(1)} &\approx&  \frac{G^2}{2}\frac{N\rho_0}{2}\int_{-\mu_5}^\infty dE\frac{1-\tilde{f}_{\beta}(E)}{E} \nonumber\\
&& \times\bar{u}_R(p_f)t^a \gamma_0 u_R(p_i)\bar{U}_+(q_f)t^a U_+(q_i) \nonumber\\
&&+ \frac{G^2}{2}\int\frac{d^3k}{(2\pi)^3} \frac{1}{\mu_5-|\vec{k}|} \nonumber\\
&& \times\bar{u}_L(p_f)t^at^b\gamma_0u_L(p_i) \bar{U}_+(q_f)t^bt^aU_+(q_i) \ , \nonumber\\
 \label{MOneCompApp}
\end{eqnarray}
which yields Eq.~(\ref{MOneComp}).

\section{Mean-field approach for Kondo effect at finite $\mu$} \label{App:finite_mu}

\begin{figure}[t!]
    \begin{minipage}[t]{1.0\columnwidth}
        \begin{center}
            \includegraphics[clip, width=1.0\columnwidth]{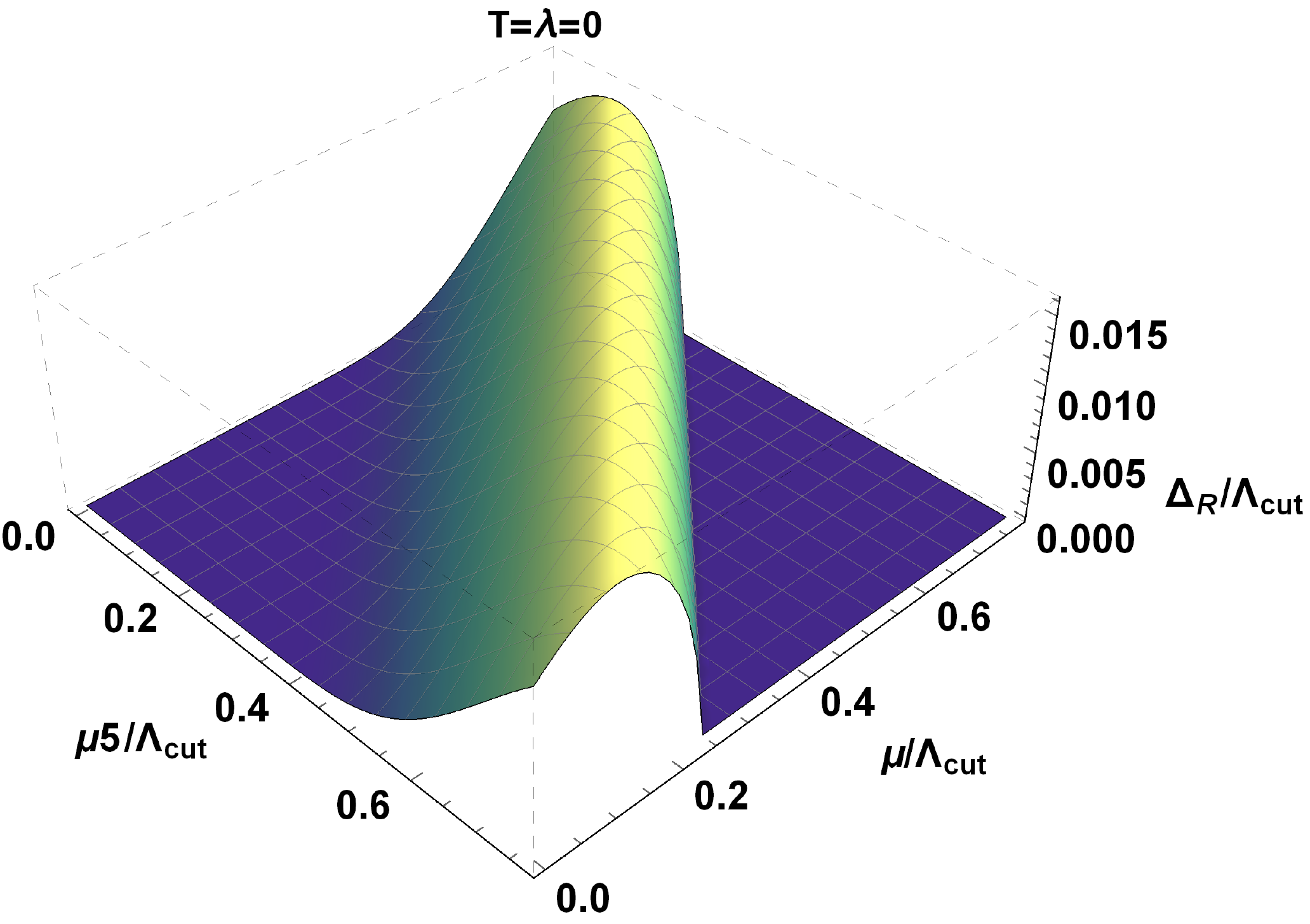}
        \end{center}
    \end{minipage}
    \vspace{5pt}
    \begin{minipage}[t]{1.0\columnwidth}
        \begin{center}
            \includegraphics[clip, width=1.0\columnwidth]{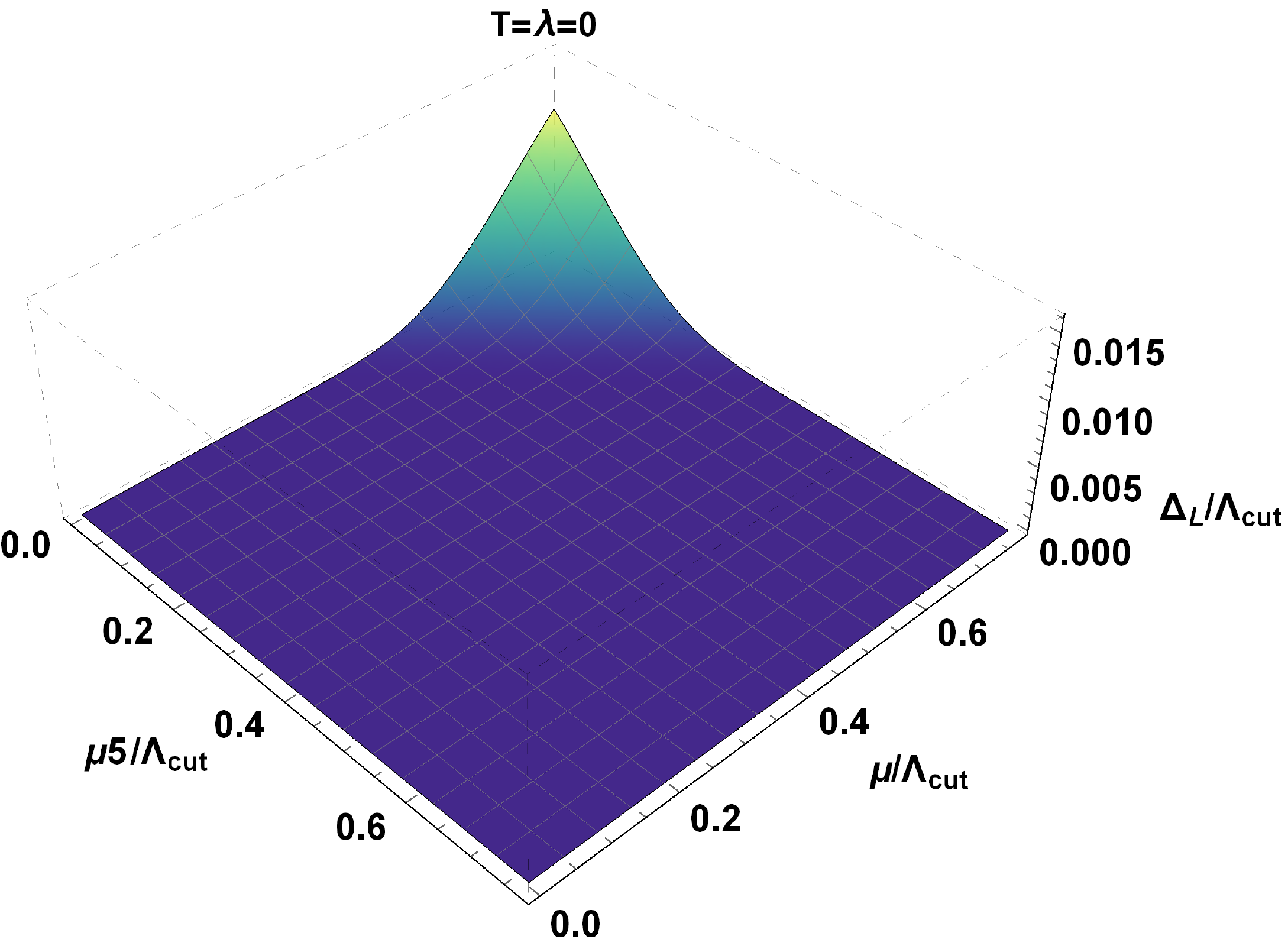}
        \end{center}
    \end{minipage}
    \caption{$\mu$-$\mu_5$ phase diagram of Kondo condensate $\Delta_{R(L)}$ at $T=\lambda=0$ using $\tilde{G}=2/\Lambda_\mathrm{cut}^2$ and $N=3$. (Top) $\Delta_{R}$. (Bottom) $\Delta_{L}$.
Vanishing $\Delta_{R}$ in the large $\mu+\mu_5$ region is a model artifact.
}
\label{fig:mu-mu5}
\end{figure}

In this appendix, in order to compare the Kondo effects at finite $\mu$ and $\mu_5$, we show the phase diagram at finite $\mu$ using the same formalism as those in the main text.
In the upper panel of Fig.~\ref{fig:mu-mu5}, we show the $\mu$-$\mu_5$ phase diagram of $\Delta_R$.
In the region with large $\mu$ and/or $\mu_5$, we find the appearance of the Kondo phase with nonzero $\Delta_R$. Note that, in the region with large $\mu+\mu_5$, $\Delta_R$ is suddenly suppressed and becomes zero, but this behavior is an artifact from the ultraviolet cutoff, as mentioned in the main text.
Therefore, we cannot conclude the true physics in this region, which is beyond the scope of this model.
As shown in the lower panel of Fig.~\ref{fig:mu-mu5}, in the region with large $\mu$ but small $\mu_5$, we find that $\Delta_L$ is also enhanced.
This behavior indicates that the ``usual" Kondo effect induced by only finite chemical potential is realized, where both the right-handed and left-handed condensates contribute to the Kondo effect (namely, $\Delta_R \approx \Delta_L$). 

\section{Order of phase transition at finite $T$} \label{App:order}

\begin{figure}[t!]
    \begin{minipage}[t]{1.0\columnwidth}
        \begin{center}
            \includegraphics[clip, width=1.0\columnwidth]{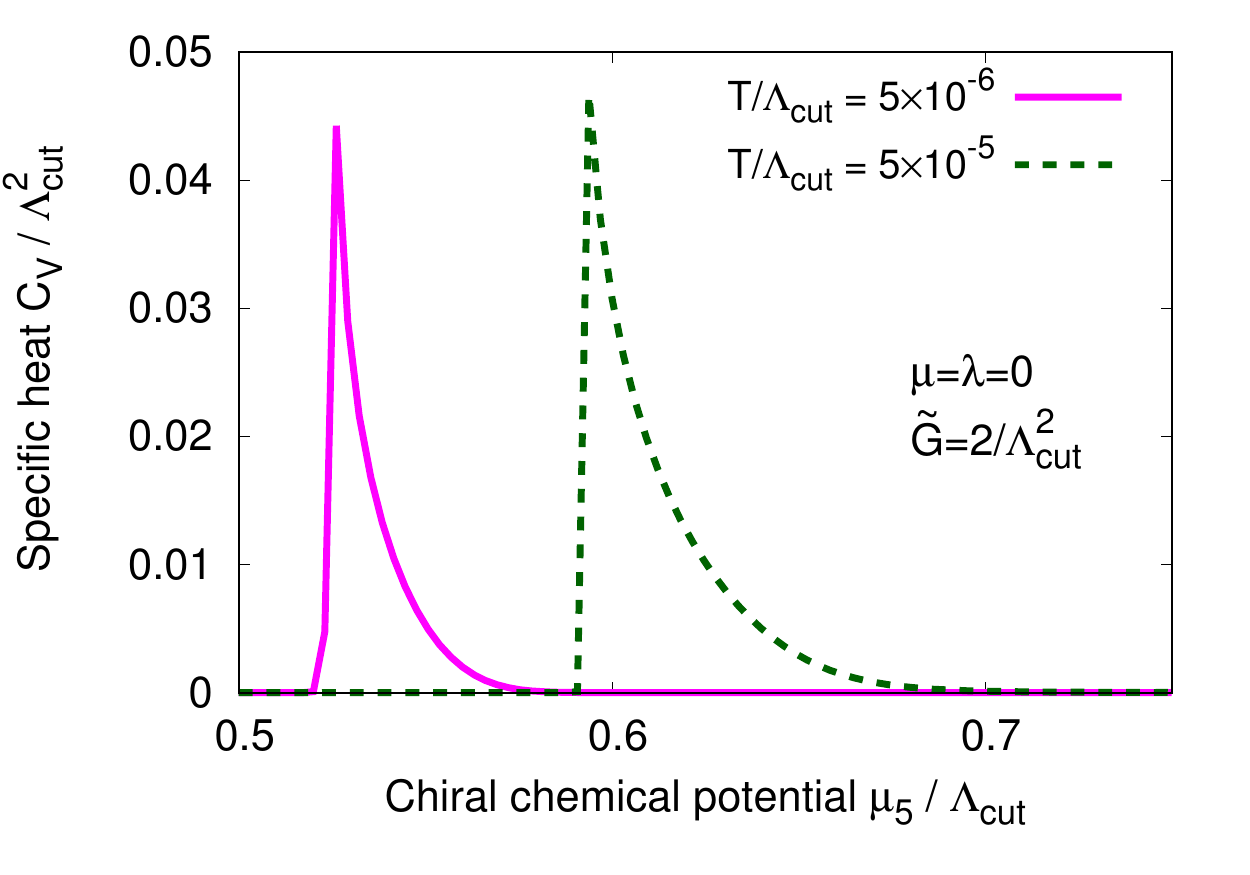}
        \end{center}
    \end{minipage}
    \caption{$\mu_5$ dependence of specific heat $C_V$ at $T/\Lambda_\mathrm{cut} = 5 \times 10^{-6}$ and $5 \times 10^{-5}$.
}
\label{fig:specifc_heat}
\end{figure}

As shown in Sec.~\ref{Sec_3-4}, at zero temperature, the value of the Kondo condensate is exponentially suppressed as $\mu_5$ decreases.
On the other hand, when a nonzero temperature is switched on, the transition is transformed into the second-order phase transition.

A second-order phase transition is characterized by the discontinuous behavior of a susceptibility (the second derivative with respect to a parameter) near the transition region.
Here, in order to check the phase transition for the Kondo condensate at finite temperature, we investigate two types of susceptibilities.

First, we investigate the specific heat defined as $C_V \equiv - T' \frac{\partial^2 }{\partial T'^2}\Omega(T',\mu_5;\Delta_{R(L)}) |_{\mu_5,T'=T}$ at a fixed temperature $T$, where the thermodynamic potential $\Omega(T,\mu_5;\Delta_{R(L)})$ is given by Eq.~(\ref{omega}).
As shown in Fig.~\ref{fig:specifc_heat}, we find a discontinuous behavior of $C_V$ between the normal phase at low $\mu_5$ and the Kondo phase at high $\mu_5$.
This discontinuity indicates that the transition at $\mu_5 \neq 0$ and $T \neq 0$ is second order. Note that the entropy density [$s \equiv -\frac{\partial}{\partial T'} \Omega(T',\mu_5;\Delta_{R(L)}) |_{\mu_5,T'=T}$] is confirmed to be a continuous function for all $\mu_5$. We also note that the discussion at finite $\mu_5$ is the same as the transition of the Kondo condensate at finite $\mu$~\cite{Yasui:2016svc,Yasui:2017izi}.

\begin{figure}[t!]
    \begin{minipage}[t]{1.0\columnwidth}
        \begin{center}
            \includegraphics[clip, width=1.0\columnwidth]{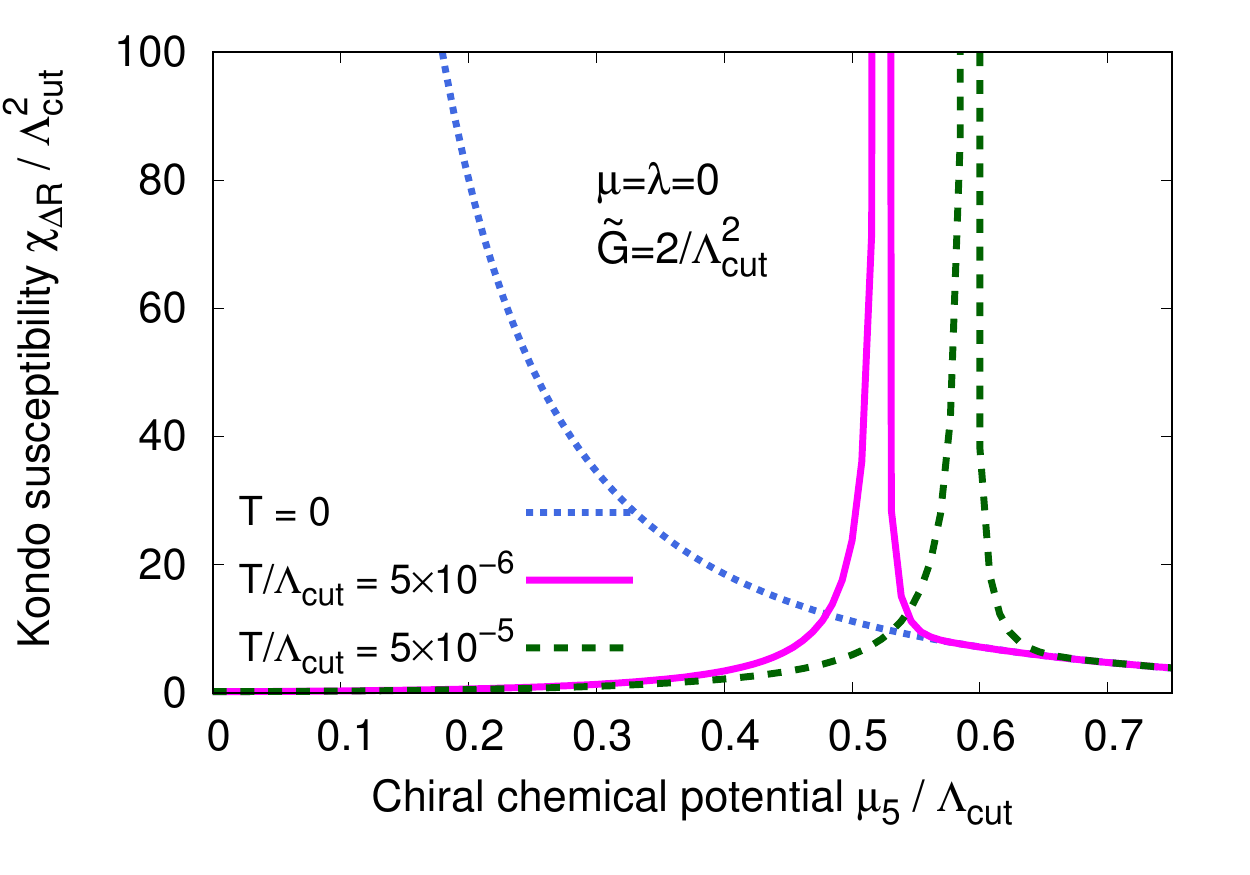}
        \end{center}
    \end{minipage}
    \caption{$\mu_5$ dependence of Kondo susceptibility $\chi_{\Delta R}$ at $T/\Lambda_\mathrm{cut} =0$, $5 \times 10^{-6}$ and $5 \times 10^{-5}$.
}
\label{fig:Kondo_Susceptibility}
\end{figure}

Second, we also investigate the {\it Kondo susceptibility} for the Kondo condensate $\Delta_R$ defined as 
\begin{eqnarray}
\chi_{\Delta R} \equiv -\frac{\partial^2}{\partial h_R^2}\tilde{\Omega}(T,\mu_5;\Delta_{R(L)})\Big|_{h_R=0} \ , 
\end{eqnarray}
where we put the minus sign in order for $\chi_{\Delta R}$ to be positive.
In this expression, $\tilde{\Omega}(T,\mu_5;\Delta_{R(L)})$ is the thermodynamic potential in the presence of an external field $h_{R(L)}$ for the Kondo condensate, which is obtained by the modified Lagrangian 
\begin{eqnarray}
\tilde{\cal L}_{\rm MF} \equiv {\cal L}_{\rm MF} + \sum_{i=R,L}\Big[h_i \bar{\Psi}_v(1+\hat{\bm p}\cdot{\bm \gamma})\psi_i + {\rm h.c.} \Big]\ ,
\end{eqnarray}
where ${\cal L}_{\rm MF}$ is defined as Eq.~(\ref{eq:MF}).

The resultant $\mu_5$ dependence of the Kondo susceptibility at zero and finite temperatures is shown in Fig.~\ref{fig:Kondo_Susceptibility}.
This figure shows that, at zero temperature, $\chi_{\Delta R}$ monotonically decreases as $\mu_5$ increases.
On the other hand, when a temperature is switched on, $\chi_{\Delta R}$ has a sharp peak at nonzero $\mu_5$, which clearly shows that the phase transition is of second order.
Note that, because it is difficult to numerically obtain the curve for $T=0$ in Fig.~\ref{fig:Kondo_Susceptibility}, we plotted an approximate analytic solution
\begin{eqnarray}
\chi_{\Delta R}|_{T=0} \approx \frac{4}{\tilde{G}}\left(\frac{\pi^2}{N\tilde{G}\mu_5^2}-1\right)\ ,
\end{eqnarray}
which is obtained under an assumption of $\Delta_R\ll \mu_5,\Lambda_{\rm cut}$ as in Eq.~(\ref{DeltaRZeroT}).

\bibliography{reference}
\end{document}